\documentclass[]{interact}
\usepackage{amssymb}
\usepackage{amsmath}
\usepackage{url}
\usepackage{algorithm}
\usepackage{changes}
\usepackage{tabularx}
\usepackage{algpseudocode}
\usepackage{tikz}
\usetikzlibrary{arrows.meta, positioning, fit, calc}

\usepackage{epstopdf}
\usepackage[caption=false]{subfig}

\usepackage[longnamesfirst,sort]{natbib}
\bibpunct[, ]{(}{)}{;}{a}{,}{,}
\renewcommand\bibfont{\fontsize{10}{12}\selectfont}

\theoremstyle{plain}

\theoremstyle{definition}

\theoremstyle{remark}

\begin{document}

\articletype{RESEARCH ARTICLE}

\title{A microservices-based endpoint monitoring platform with predictive NLP models for real-time security and hate-speech risk alerting}

\author{
\name{Darlan Noetzold\textsuperscript{a,c}\thanks{CONTACT Darlan Noetzold. Email: darlan.noetzold@gmail.com}, 
Anubis Graciela De Moraes Rossetto\textsuperscript{b},
Juan Francisco De Paz Santana\textsuperscript{c}, 
Valderi Reis Quietinho Leithardt\textsuperscript{a,c}}
\affil{\textsuperscript{a}Expert Systems and Applications Laboratory (ESALAB), Faculty of Science, University of Salamanca, Salamanca, Spain; \\
\textsuperscript{b}Federal Institute Sul-Rio-grandense (IFSUL), Passo Fundo, Brazil; \\
\textsuperscript{c}Instituto Universitário de Lisboa (ISCTE-IUL), ISTAR, Lisboa, 1649-026, Portugal}
}



\maketitle

\begin{abstract}
Organizations increasingly depend on endpoint devices and corporate communication channels, yet they still face critical risks such as sensitive data leakage, suspicious user behavior, and the circulation of hateful or harmful language in workplace contexts. Current solutions frequently address these issues in isolation (e.g., productivity tracking, data loss prevention, or hate-speech detection), limiting correlation across signals and delaying incident response. This work proposes a unified, microservices-based platform that collects endpoint telemetry and applies predictive natural language processing models to support real-time security and compliance alerting. The architecture is modular and scalable, relying on RabbitMQ for event ingestion and routing and Redis for low-latency data access and alert delivery. For text classification, transformer-based models such as BERT are evaluated for hate-speech risk detection, achieving an average accuracy of 87\%. Experimental results indicate that the proposed platform can promptly surface indicators of data exfiltration and policy violations while centralizing alert management, providing an integrated framework that combines monitoring, security analytics, and predictive capabilities.
\end{abstract}


\section*{Introduction}

Organizations currently face a complex landscape of challenges regarding cybersecurity, operational efficiency, and workplace well-being \citep{leithardt12014privacy, 11503355}. In this context, the unauthorized exposure of sensitive information and the dissemination of harmful content have emerged as critical risks with profound ramifications for both institutional stability and employee health. The imperative to address these issues is undeniable, as the absence of proactive preventive strategies can lead to substantial financial losses, severe reputational damage, and psychological effects on the workforce \citep{dafhalla2025innovative}.

Hate speech, in its various manifestations, represents a harmful phenomenon that permeates multiple organizational spheres \citep{reshma2025dynamic}. With its proliferation across digital platforms, awareness of this problem has grown significantly among researchers and policymakers \citep{emon2025integration, A1}. A notable example of the institutional impact is the legal precedent involving Tesla, where the company was ordered to pay over \$130 million in damages to a former employee due to a racially hostile work environment. The court's decision underscored the organization's failure to monitor internal conduct and protect its staff from emotional and psychological harm, highlighting the legal and ethical necessity of oversight \citep{leithardt12014privacy, NOETZOLD2026131705}.

Furthermore, hate crimes and discriminatory behaviors have shown a growing presence within organizational structures, as documented by the US Department of Justice \citep{zhang2021hate}. Consequently, there is an urgent need for constant monitoring systems capable of mitigating hate speech and preserving a safe, productive environment \citep{doi:10.1080/17512786.2023.2288918, 11488882}. Parallel to this, data leakage remains a crucial challenge, with the average cost of a data breach reaching US\$4.35 million in recent years \citep{ibm}. The rise of remote work has further intensified these risks, requiring innovative approaches to ensure the integrity of business systems and sensitive data.

In response to these challenges, Electronic Performance Monitoring (EPM) has emerged as a practice for collecting and analyzing data related to employee activities through digital technologies. EPM encompasses the systematic capture of information regarding actions, interactions, and results, allowing for objective, data-driven evaluations \citep{kalischko2021electronic, 5503914}. While studies address the ethical and legal concerns of EPM, emphasizing the balance between organizational needs and employee rights \citep{sherif2021empowering, wieser2023employers}, there remains a gap in integrated solutions that combine security telemetry with behavioral analysis.

The primary contribution of this work lies in the proposal of a unified, microservices-based framework that integrates endpoint monitoring techniques with predictive natural language processing (NLP) models. Unlike fragmented solutions, this research contributes: (i) a modular architecture designed for high-volume data ingestion and real-time alert orchestration; (ii) the integration of multi-source telemetry, including keystroke analysis, network traffic monitoring, and process auditing, into a single security pipeline; and (iii) the application of transformer-based machine learning models to detect hate speech risk indicators with high accuracy. By centralizing these capabilities, the study provides a scientifically-grounded approach to simultaneously addressing data loss prevention and workplace compliance.

The remainder of this article is organized as follows: Section II describes the fundamental concepts and technologies supporting the solution; Section III reviews related works and identifies existing gaps; Section IV details the proposed microservices architecture and component integration; Section V focuses on the implementation of monitoring techniques and the predictive models; Section VI presents the experimental evaluation and performance results; and finally, Section VII provides concluding remarks and suggestions for future research.

\section{Related Works} \label{sec_related_works}

To realize the proposed solution, preliminary research was conducted into studies that address the problem in question as well as closely related scenarios. The literature spans workplace/endpoint monitoring solutions, predictive models for hate speech detection, and continuous monitoring approaches for security and privacy-aware analytics. The following discussion reviews representative contributions and supports a direct comparison with the proposed application.

Companies, developers, and researchers have explored the potential of workplace monitoring tools. Commercial solutions such as Kickidler \citep{Kickidler}, ActivTrak \citep{activtrak}, FSense \citep{fsense}, DeskTime \citep{DeskTime1}, and StaffCop \citep{staffcop1} provide functionalities like real-time screen monitoring, keylogging, and productivity reports. However, these solutions are primarily designed for productivity monitoring and do not integrate hate speech detection with vulnerability-oriented checks. Furthermore, most commercial solutions have limitations in their free versions, such as restrictions on the number of computers that can be monitored.

Research efforts in harmful-content detection include systems and models focused on aggression and hate speech. Modha et al. \citep{MODHA2020113725} presented an approach for detecting and visualizing aggression on social media through a browser plug-in that highlights aggressive comments. Paschalides et al. \citep{paschalides2020mandola} introduced Mandola, a system for reporting and monitoring online hate speech based on an ensemble classifier and a modular processing pipeline. Roy et al. \citep{turn0search0} proposed a cross-modal alignment approach for hate speech detection with uncertainty-aware mechanisms. Smith et al. \citep{turn0search2} introduced PREDICT, a multi-agent debate simulation framework for generalized hate speech detection. A multimodal strategy for hate speech event detection integrating text and images was presented in \citep{turn0search4}, while Ganguly et al. \citep{turn0academia23} explored transformer ensembles for multimodal hate speech detection. For Brazilian Portuguese, Oliveira et al. \citep{oliveira2024tupy} introduced TuPy-E and compared learning approaches, highlighting the strong performance of BERT-style models. Forzinetti et al. \citep{forzinetti2024indicators} examined indicators for characterizing hate speech and the alignment of automated detection with expert assessment in multilingual contexts. Complementarily, Bharathi et al. \citep{10717313} proposed InsightGuard, combining machine learning and computer vision to monitor screen usage and ergonomic factors with an alert mechanism.

Additional studies address foundational aspects for large-scale monitoring, such as data processing, privacy, and security analytics. Dafhalla et al. \citep{dafhalla2025innovative} discussed analytical solutions across IoT and cloud-continuum environments, emphasizing enhanced data processing. Reshma et al. \citep{reshma2025dynamic} introduced a differential privacy framework for heterogeneous data scenarios. Gangwani et al. \citep{Gangwani2026} evaluated convolutional autoencoders for anomaly detection on control-flow representations, and Ababneh et al. \citep{Ababneh2026} proposed a hybrid intrusion detection and ensemble learning system for continuous monitoring.

The current study presents a novel microservices-based solution that integrates endpoint monitoring techniques with predictive models for organizational environments. As summarized in Table \ref{relatedworks}, this work combines keylogging, screen capture, process monitoring, internet traffic monitoring, vulnerability-oriented checks, and predictive hate speech detection within a unified framework, supporting real-time analysis and alerting. In addition, the proposed framework evaluates transformer-based models for hate speech classification, achieving an accuracy of approximately 87\%. This combination addresses a gap in the literature by jointly tackling cybersecurity-relevant monitoring and harmful-content risk detection in a single architecture designed for organizational use.

\begin{table}[h]
\caption{Comparison of related works (rows) by capabilities (columns)}
\label{relatedworks}
\centering
\resizebox{\textwidth}{!}{%
\begin{tabular}{p{4.4cm} c c c c c c c c c}
\hline
Work & Key Capture & Screen Capture & Process Monitoring & Internet Traffic & Vulnerability Scan & Hate Speech Alert & ML Integration & Multimodal & Dashboard \\
\hline
Kickidler \citep{Kickidler} & X & X & X & - & - & - & - & - & X \\
ActivTrak \citep{activtrak} & X & X & - & - & - & - & - & - & X \\
FSense \citep{fsense} & - & - & - & - & - & - & - & - & X \\
DeskTime \citep{DeskTime1} & - & - & - & - & - & - & - & - & X \\
StaffCop \citep{staffcop1} & X & X & X & X & - & - & - & - & X \\
Paschalides et al. \citep{paschalides2020mandola} & - & - & - & X & - & X & X & - & - \\
Modha et al. \citep{MODHA2020113725} & - & - & - & - & - & X & - & - & - \\
Roy et al. \citep{turn0search0} & - & - & - & - & - & X & X & X & - \\
Smith et al. \citep{turn0search2} & - & - & - & - & - & X & X & X & - \\
Johnson et al. \citep{turn0search4} & - & - & - & - & - & X & X & X & - \\
Ganguly et al. \citep{turn0academia23} & - & - & - & - & - & X & X & X & - \\
Oliveira et al. \citep{oliveira2024tupy} & X & - & - & - & - & X & X & - & X \\
Forzinetti et al. \citep{forzinetti2024indicators} & - & - & - & - & - & X & X & X & - \\
Bharathi et al. \citep{10717313} & - & X & - & - & - & - & X & - & X \\
Dafhalla et al. \citep{dafhalla2025innovative} & - & - & - & - & - & - & X & - & - \\
Reshma et al. \citep{reshma2025dynamic} & - & - & - & - & - & - & - & - & - \\
Gangwani et al. \citep{Gangwani2026} & - & - & - & - & - & - & X & - & - \\
Ababneh et al. \citep{Ababneh2026} & - & - & - & - & - & - & X & - & X \\
\textbf{Proposed} & \textbf{X} & \textbf{X} & \textbf{X} & \textbf{X} & \textbf{X} & \textbf{X} & \textbf{X} & \textbf{X} & \textbf{X} \\
\hline
\end{tabular}}
\end{table}

\section{Methodology}\label{sec_methodology}

This section describes the methodological foundations and the technical design adopted to build, integrate, and operate the proposed monitoring and content-risk detection framework. The methodology follows a requirements-driven engineering approach, in which functional and non-functional requirements (e.g., latency, scalability, maintainability, and auditability) guided the decomposition of the solution into independent services and well-defined interfaces. From a research perspective, the proposed system is treated as an artifact whose design choices are justified by the need to (i) collect endpoint activity signals, (ii) transform such signals into structured evidence, and (iii) support timely decision-making through alert generation and predictive classification.

The overall method is organized into four stages. First, requirements elicitation and threat-oriented analysis were conducted to identify the monitored signals and define alert criteria (e.g., policy violations, suspicious processes, risky network destinations, and offensive textual content). Second, an architectural design stage specified the logical layers of the system (endpoint collection, ingestion/orchestration, storage, analytics/classification, and presentation), as well as the data contracts connecting them. Third, the implementation stage instantiated each module and defined the execution flows for (a) policy synchronization, (b) alert creation, enrichment, and persistence, and (c) hate speech classification requests and responses. Finally, the evaluation stage assessed correctness and performance, focusing on the ability to detect events promptly and to provide actionable context for incident analysis.

The methodology also accounts for operational and governance aspects intrinsic to organizational monitoring. In particular, the design assumes that data collection and analysis must be performed under explicit authorization and according to internal policies and applicable regulations. To that end, the framework incorporates mechanisms that support traceability (structured alert records, evidence capture, and reproducible classification outputs) and controlled access (role-based visibility of alerts and configuration artifacts). The next subsections present the endpoint monitoring component, the service orchestration layer, the user-facing interface, and the predictive models and classification pipeline.

\begin{figure}[ht]
\centering
\caption{Detailed architecture and data flows of the proposed framework.}
\label{fig:full_architecture}
\resizebox{\textwidth}{!}{%
\begin{tikzpicture}[
  font=\normalsize, 
  node distance=12mm and 25mm,
  box/.style={draw, rounded corners, align=center, inner sep=8pt, minimum height=12mm, minimum width=52mm},
  svc/.style={box, fill=gray!10},
  data/.style={box, fill=gray!5},
  agent/.style={box, fill=gray!15},
  arrow/.style={-Latex, thick},
  lab/.style={midway, fill=white, inner sep=1.5pt, font=\scriptsize}
]

\node[agent] (endpoint) {Employee Endpoint\\(Monitoring Agent)};
\node[agent, below=12mm of endpoint] (threads) {Parallel collectors\\(sniffer / key capture / etc.)};
\node[agent, below=12mm of threads] (evidence) {Alert Evidence\\(metadata + context)};

\node[svc, right=40mm of endpoint] (api) {Central API Layer\\(ingestion \& orchestration)};
\node[svc, below=10mm of api] (auth) {Auth \& User Service};
\node[svc, below=10mm of auth] (policy) {Policy Service};
\node[svc, below=10mm of policy] (alerts) {Alert Service};
\node[svc, below=10mm of alerts] (modelgw) {Prediction API};

\node[svc, right=40mm of api] (ui) {Front-End Application};
\node[data, below=10mm of ui] (db) {Relational Database};
\node[data, below=10mm of db] (blob) {Evidence Storage};
\node[data, below=10mm of blob] (cache) {Cache Layer};
\node[data, below=10mm of cache] (queue) {Async Messaging};
\node[data, below=10mm of queue] (obs) {Observability};

\node[draw, dashed, rounded corners, fit=(endpoint) (threads) (evidence), inner sep=15pt, label={[font=\bfseries\large]above:Endpoint Layer}] (group1) {};
\node[draw, dashed, rounded corners, fit=(api) (auth) (policy) (alerts) (modelgw), inner sep=20pt, label={[font=\bfseries\large]above:Service Layer}] (group2) {};
\node[draw, dashed, rounded corners, fit=(ui) (db) (blob) (cache) (queue) (obs), inner sep=20pt, label={[font=\bfseries\large]above:Data / UI / Ops}] (group3) {};


\draw[arrow] (endpoint) -- node[lab] {(A)} (threads);
\draw[arrow] (threads) -- node[lab] {(B)} (evidence);

\draw[arrow] (policy.west) -- ++(-12mm,0) |- node[lab, pos=0.25] {(1) policy} (endpoint.east);

\draw[arrow] (evidence.east) -- ++(12mm,0) |- node[lab, pos=0.75] {(2) alert event} (api.west);

\draw[arrow] (ui.north) -- ++(0,8mm) -| node[lab, pos=0.25] {(9) queries} (api.north);

\draw[arrow] (api.east) -- ++(10mm,0) |- node[lab, pos=0.8] {} (db.west);
\draw[arrow] (api.east) -- ++(15mm,0) |- node[lab, pos=0.8] {(4) store} (blob.west);
\draw[arrow] (api.east) -- ++(20mm,0) |- node[lab, pos=0.8] {(13) cache} (cache.west);
\draw[arrow] (api.east) -- ++(25mm,0) |- node[lab, pos=0.5] {(14) telemetry} (obs.west);

\draw[arrow] (api.west) ++(0,-4mm) -- ++(-8mm,0) |- node[lab, pos=0.4] {(7) classify} (modelgw.west);

\draw[arrow] (modelgw.east) -- ++(8mm,0) |- node[lab, pos=0.75] {} (db.west);

\draw[arrow] (alerts.east) -- ++(12mm,0) |- node[lab, pos=0.75] {(6) async tasks} (queue.west);

\draw[arrow] (auth.east) -- node[lab, pos=0.5] {} (db.west);
\draw[arrow] (policy.east) -- node[lab, pos=0.5] {(11)} (db.west);

\draw[arrow] (api) -- node[lab] {(5)} (auth);
\draw[arrow] (auth) -- (policy);
\draw[arrow] (policy) -- (alerts);
\draw[arrow] (alerts) -- (modelgw);

\end{tikzpicture}%
}
\end{figure}
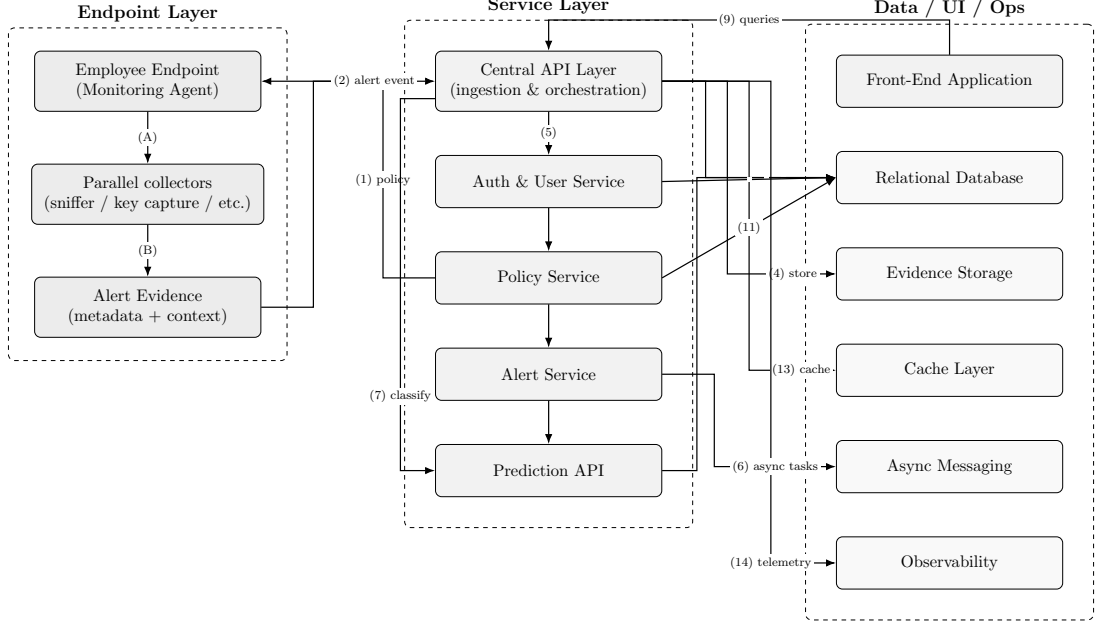

\subsection{Spyware}
\begin{figure}[ht]
\centering
\caption{Alert generation flow.}
\label{fig:generate_alert}
\includegraphics[scale=0.5]{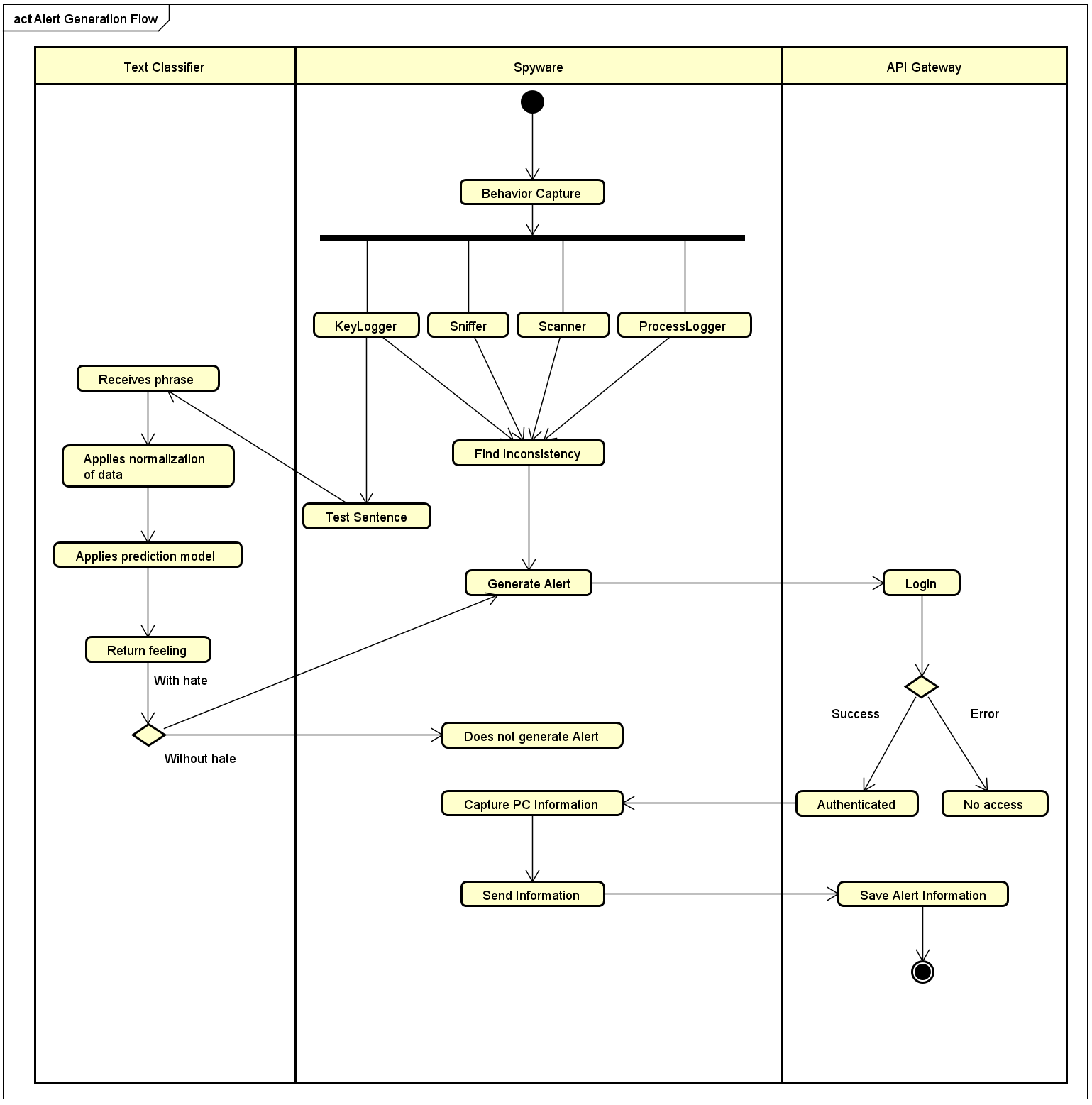}%
\end{figure}

The diagram provided in Figure \ref{fig:generate_alert} is a high-level overview designed to illustrate the execution flow of the alert generation process within the Spyware component. Its purpose is to provide a simplified visualization of the system's operation rather than detailing every individual component and process. While this diagram focuses on the general execution flow, the following subsections explain the specific components and criteria used for generating alerts.

The Spyware developed is a Python script compiled in executable format using the Freeze library. This script monitors various metrics, capturing the data needed to create alerts. The monitoring and data captured were defined through the process of planning and selecting the application's functional requirements.

The diagram in Figure \ref{fig:generate_alert} shows the alert generation flow, with the initial flow being the capture of user behavior. The script is divided into several \emph{Threads}, each with specific functions that are described below.

\begin{itemize}
    \item \textbf{Sniffer}: Monitors network packet traffic in real-time, capturing relevant data. It operates in a separate \emph{Thread} and has attributes such as log lists, sets for captured DNS queries and blocked sites, and counters for captured packets. The main function, \texttt{sniffer()}, checks DNS packets and generates alerts if it finds queries to blocked sites. The list of blocked sites is dynamically updated via the \texttt{update\_aux\_data()} function, which retrieves this data from administrator-defined blacklists maintained in the Front-End application. Packet capture is carried out in the \emph{Thread} \texttt{run()} with the Scapy library;
    
    \item \textbf{KeyLogger}: Records the keys pressed by the user. Initializes an empty log and uses the method \texttt{callback()} to handle keyboard events, recording the keys pressed. Detects special key presses, such as ''enter'', ''backspace'' and ''caps lock''. The \texttt{report()} method analyzes the log, identifies unwanted content, and sends alerts if necessary. The \emph{Thread} is started with the \texttt{start()} method;
    
    \item \textbf{Scanner}: Performs scans to find open ports on a specific target. It runs in a separate \emph{Thread} and uses the Socket library to establish connections and collect service banners on open ports. Ports and banners are stored in lists, and banners are compared with a dynamically updated list of known vulnerabilities. This list is obtained from public vulnerability databases, such as the NVD (National Vulnerability Database), and configured by the administrator for relevance to the monitored environment. Alerts are generated if matches are found;
    
    \item \textbf{ProcessLogger}: Identifies and deals with malicious processes running on the system. The \texttt{are\_malicious\_ }\texttt{process(self)} function checks for malicious processes by comparing names with a list provided by the administrator via the local API. This list is based on known threats and organizational policies. If a match is found, the process is terminated and an alert is logged. The \texttt{get\_process()} function obtains an up-to-date list of running processes.
\end{itemize}

These \emph{Threads} use specific strategies to monitor different points. From this detailed explanation, it is possible to discuss the data captured by Spyware.
Another important point is how the data that forms the Alert is captured. A script has been developed to collect essential data, including running processes, the MAC address of the monitored computer, an image of the screen at the time of the alert, and the reason for the alert.

The running processes are obtained by using the \texttt{get\_} \texttt{process()} function, while the MAC address is obtained using Python's \texttt{get\_mac\_address} library. The reason for the alert derives from the monitoring previously implemented, which checks for suspicious activity such as access to blocked sites, malicious DNS queries, and unwanted keywords.
Capturing the screen image during the alert uses the "ImageGrab" library, which encodes the image in base64 for later storage and analysis.
These strategies are essential for a complete analysis of alerts, providing context about suspicious activity on the monitored system. Implementation must follow appropriate security policies and permissions, ensuring compliance with standards and regulations.

To improve reproducibility and facilitate analysis, the alert decision can be formalized as a multi-signal detection problem. Let the monitoring stage produce binary indicators for each evidence source at time $t$:
\begin{equation}
x_i(t) \in \{0,1\}, \quad i \in \{\text{dns},\text{kw},\text{proc},\text{port},\text{hs}\},
\end{equation}
where $x_{\text{dns}}$ denotes a match with a blocked destination, $x_{\text{kw}}$ the occurrence of prohibited keywords, $x_{\text{proc}}$ a match with a flagged process, $x_{\text{port}}$ a match with a vulnerable service/banner, and $x_{\text{hs}}$ a positive hate-speech classification.

A simple decision rule is defined as:
\begin{equation}
D(t) = \mathbb{I}\left(\bigvee_i x_i(t) = 1\right),
\label{eq:alert_or_rule}
\end{equation}
where $\mathbb{I}(\cdot)$ is the indicator function. Additionally, a risk score can be computed to rank alerts by severity:
\begin{equation}
S(t) = \sum_i w_i \, x_i(t), \quad w_i \ge 0,
\label{eq:alert_score}
\end{equation}
triggering an alert when $S(t) \ge \tau$ for a configurable threshold $\tau$.

During monitoring and alert generation, the script analyzes points such as DNS, processes, open ports, and typed words, comparing them with lists containing prohibited items defined by the administrator in the Front-End application. The data is updated using the \texttt{update\_aux\_data()} function. This function obtains data from categories such as inappropriate words, vulnerable ports, malicious processes, and blocked sites via requests to endpoints from a local API. The function performs a login to obtain API access credentials and makes requests to each specific endpoint using these credentials. The data returned is stored in text files separated by semicolons. The function records log messages in case of errors during the process. Algorithm~\ref{alg:spyware_alert_generation} summarizes the end-to-end procedure for generating, enriching, and dispatching alerts, integrating multi-thread monitoring, policy synchronization, and text classification when applicable.

\begin{algorithm*}[h!]
\caption{Spyware alert generation and enrichment}\label{alg:spyware_alert_generation}
\scriptsize
\begin{algorithmic}[1]
\Require Policy lists $\mathcal{B}_{dns}, \mathcal{B}_{kw}, \mathcal{B}_{proc}, \mathcal{B}_{port}$; weights $\{w_i\}$; threshold $\tau$; update interval $\Delta$.
\Ensure Alerts sent to the central ingestion endpoint when $S(t)\ge\tau$.
\State Authenticate and call \texttt{update\_aux\_data()} to fetch $\mathcal{B}_{dns}, \mathcal{B}_{kw}, \mathcal{B}_{proc}, \mathcal{B}_{port}$.
\State Spawn monitoring threads: network, keystroke, process, port scan.
\Loop
    \If{time since last update $\ge \Delta$}
        \State Refresh policy lists via \texttt{update\_aux\_data()}.
    \EndIf
    \State Collect events from threads (e.g., DNS query, phrase, process, open port/banner).
    \State Compute indicators $x_i(t)$ for the current event using the corresponding blacklist/rule.
    \If{event contains a typed phrase}
        \State Request text classification from the Prediction Models API Gateway and set $x_{\text{hs}}(t)$ accordingly.
    \EndIf
    \State Compute score $S(t) \gets \sum_i w_i x_i(t)$.
    \If{$S(t) \ge \tau$}
        \State Capture context: process list, MAC address, timestamp, and screen evidence.
        \State Build alert payload $A \gets \langle \text{pcId}, t, \text{reason}, \text{evidence}, S(t)\rangle$.
        \State Send $A$ to the Central API Layer for storage and further processing.
    \EndIf
\EndLoop
\end{algorithmic}
\end{algorithm*}

This function is used for obtaining and updating auxiliary data used in analysis, detection, and system decisions. After configuring blacklists, implementing monitoring, and capturing the necessary data, hate speech detection is tackled. KeyLogger captures typed phrases, which are sent by HTTP request via POST to a Prediction Models API Gateway. If the phrase is classified as hate speech, an alert is generated.
This integrates hate speech detection with spyware, allowing for the identification and response to offensive content. The approach involves an external API for classification, ensuring accurate and up-to-date analysis.

\subsection{API Gateway Central}

The API Gateway Central was developed to ensure modularization and efficient data management, coordinating communication between various system components such as image storage, alerts, users, and blacklists. This architecture focuses on managing data flow and service orchestration while maintaining consistent performance under high request loads. Docker was employed to containerize the API Gateway, isolating the web service and its dependencies to simplify deployment and version control. Persistent data, including user information, alerts, and images, is managed using PostgreSQL, ensuring reliability, ACID compliance, and efficient handling of relational data.

Redis was integrated to optimize performance by caching frequently accessed data, such as queries to the \texttt{/alert} endpoint, reducing response times and load on the PostgreSQL database. Cached data is temporarily stored in memory for 20 seconds to maintain consistency without overburdening the main database \citep{Begnum:2020, Gao:2018}. RabbitMQ facilitates asynchronous communication between distributed services, enabling non-blocking task management and improving scalability. It supports the task queuing mechanism critical for workflows like alert processing, where timely and reliable communication is essential \citep{sengupta2017performance, sarwar2019scalability, ahmed2020iot}.

For the asynchronous processing path, basic queueing relations help characterize system behavior under load. Using Little's Law, the average number of pending messages $L$ in a stable queue can be related to arrival rate $\lambda$ and average waiting time $W$:
\begin{equation}
L = \lambda W.
\label{eq:littles_law}
\end{equation}
This relation is useful to guide capacity planning and to interpret observed delays during peak workloads. From an analytical standpoint, the caching effectiveness can be expressed through the cache hit ratio:
\begin{equation}
H = \frac{N_{\text{hit}}}{N_{\text{hit}} + N_{\text{miss}}},
\label{eq:cache_hit_ratio}
\end{equation}
where $N_{\text{hit}}$ and $N_{\text{miss}}$ denote the number of cache hits and misses in a given observation window. Considering a TTL-based cache with $\mathrm{TTL}=20$s, the expected response time can be approximated by:
\begin{equation}
\mathbb{E}[T] \approx H \cdot T_{\text{cache}} + (1-H)\cdot T_{\text{db}},
\label{eq:expected_latency}
\end{equation}
where $T_{\text{cache}}$ and $T_{\text{db}}$ represent the service times when the request is served from cache or from the database, respectively.

FlywayDB was utilized for automated database migrations, allowing consistent version-controlled schema updates across different environments. This ensures that changes are applied reliably during deployment and supports collaboration by tracking schema modifications. The observability stack integrates Spring Boot Actuator, Prometheus, and Grafana to monitor system metrics such as CPU usage, memory consumption, and request throughput. Prometheus collects real-time metrics from the system, while Grafana visualizes them in dashboards for quick identification of bottlenecks. Logs are generated using SLF4J, supporting comprehensive diagnostics and operational monitoring.

This architecture is designed to handle high workloads while providing modularity, scalability, and reliability. By leveraging containerization, caching, and asynchronous communication, the system ensures efficient data flow and consistent performance across all components. Observability tools enhance monitoring and visualization, enabling administrators to maintain system health and address performance issues proactively.

\subsubsection{Performance Improvement Measures}

Several structural and development measures were applied to improve the performance and agility of the application, resulting in a more responsive experience. These optimizations target different layers of the system stack, from application initialization and runtime configuration to database indexing and observability overhead. Each measure was selected based on profiling results and benchmarking studies that identified bottlenecks in startup time, request throughput, and resource consumption. The following measures describe the technical rationale and the observed impacts on the system's efficiency:

\begin{itemize}
\item The application has been configured to start up in lazy mode, loading only essential components during the initialization phase. In traditional eager initialization, all beans, services, and dependencies are instantiated at startup, which can significantly increase boot time, especially in microservice architectures with numerous auto-configured components. By deferring the instantiation of non-critical beans until they are first accessed, this strategy reduces the initial memory footprint and accelerates deployment cycles. Empirical tests showed a reduction in startup time of approximately 30--40\% compared to the default eager mode, enabling faster recovery from failures and more efficient resource utilization during horizontal scaling events.

\item Disabling unnecessary auto-configurations focuses the application, saving resources and simplifying maintenance. Spring Boot's default mechanism simplifies development by automatically configuring beans based on classpath dependencies, but this convenience comes at the cost of loading numerous configuration classes that may not be relevant to the application's specific runtime requirements. By explicitly excluding components related to unused messaging systems or batch processing, the application reduces the number of beans in the application context and shortens the configuration evaluation phase. Profiling indicated a reduction in context initialization time by approximately 15--20\%, contributing to overall system responsiveness.

\item Migrating the servlet container from Apache Tomcat to Undertow improved the overall performance of the application, making it more agile and responsive under high concurrency. Undertow is a lightweight, non-blocking servlet container that uses a more efficient threading model and lower memory overhead per connection compared to traditional containers. Benchmark tests demonstrated a 10--15\% improvement in request throughput and a reduction in average response latency, particularly for endpoints with high request rates. This change allows the system to handle a higher volume of concurrent users without proportional increases in CPU and memory consumption \citep{smith2022comparative}.

\item Disabling Java Management Extensions (JMX) avoids consuming extra resources, given the existence of other metrics tools within the observability stack. While JMX provides a standardized mechanism for managing Java applications, it introduces runtime overhead due to the registration of MBeans and periodic metrics collection. Since the system already integrates Prometheus and Grafana for real-time monitoring, JMX was deemed redundant. This optimization resulted in a measurable reduction in CPU usage and simplified the runtime configuration by removing an additional management interface that would otherwise require security hardening.

\item Controlling Hibernate and JPA logs reduced computational cost and improved processing efficiency. These frameworks generate extensive logging output by default, including SQL statements and transaction lifecycle events, which imposes a significant I/O burden in production environments. By configuring logging levels to capture only warnings and errors, the system decreased the volume of log data that must be ingested and analyzed. Profiling indicated that excessive logging contributed to a measurable portion of CPU time in database-intensive workloads, and this reduction allowed for more predictable performance during peak traffic.

\item Creating sequential indexes optimized storage efficiency and improved query performance by ensuring that newly inserted rows are clustered together on disk. Traditional random index generation can lead to fragmentation and suboptimal page utilization in database engines, especially under high insert rates. Sequential index generation reduces the number of page splits, improves cache locality, and accelerates range scans and index lookups. Benchmarking showed a 10--12\% improvement in insert throughput and a reduction in index fragmentation over time, ensuring stable query performance as the dataset grows.
\end{itemize}

These improvements have resulted in promising performance and scalability results, allowing the application to meet user demands more effectively. To validate the cumulative effect, load tests were conducted comparing the baseline configuration against the optimized version. The results showed that startup time was reduced from an average of 18 seconds to 11 seconds, while request throughput increased by 15\% under sustained load. Furthermore, average response latency decreased by approximately 15\% at the 95th percentile, and the steady-state memory footprint was reduced by 12\%, supporting higher container density and better cost efficiency in the deployment environment.

\subsection{Front-End Application}

A Front-End application was developed to provide a user-friendly interface for viewing and managing the system's data, including functionalities such as viewing alerts, updating blacklists, and administering other aspects of the solution. The interface ensures ease of navigation and interaction, catering to both administrators and regular users through a responsive design that adapts to different screen sizes and devices. The application was built using modern web technologies, including HTML5, CSS3, and JavaScript frameworks, to deliver a seamless user experience with minimal latency and intuitive workflows.

The application integrates seamlessly with the backend through RESTful API endpoints, maintaining consistency through synchronized development and iterative improvements based on user feedback and business requirements. Communication between the front-end and the Central API Layer follows a stateless request-response model, where authentication tokens are included in HTTP headers to ensure secure and traceable interactions. The front-end employs asynchronous data fetching techniques to prevent blocking operations, allowing users to continue interacting with the interface while data is being retrieved or submitted in the background.

Role-based authentication is a core feature of the application, controlling access to sensitive functionalities and data. Administrators can manage system-wide alerts and blacklists, configure monitoring policies, and oversee user accounts, while regular users are restricted to viewing and logging their own alerts. This approach ensures secure access control and segregation of responsibilities, reducing the risk of unauthorized modifications and supporting compliance with organizational security policies. User sessions are managed through secure cookies or token-based mechanisms, with automatic expiration and renewal to balance security and usability.

Key functionalities include paginated views of alerts for both administrators and logged-in users, reducing the amount of data transferred per request and improving rendering performance. Search capabilities allow users to filter alerts by multiple criteria, including MAC address, timestamp, alert type, and severity score, enabling rapid identification of relevant events. Pages dedicated to updating blacklists provide administrators with the ability to add, modify, or remove entries for blocked domains, prohibited keywords, malicious processes, and vulnerable service banners. These changes are immediately synchronized with the backend and propagated to all active monitoring agents through the policy synchronization mechanism described in Section~\ref{sec_methodology}.

User registration and login mechanisms enhance access security, with distinct roles granting different permissions based on organizational hierarchy and operational needs. Password policies enforce minimum complexity requirements, and failed login attempts are logged and rate-limited to mitigate brute-force attacks. Additionally, the system enables manual registration of alerts through JSON downloads, allowing users to export alert records for offline analysis, archival, or submission as evidence in incident response workflows. Each exported JSON file includes metadata such as the alert identifier, timestamp, reason, associated evidence, and a cryptographic hash to ensure data integrity and authenticity.

The front-end application also incorporates real-time notification features, where users receive visual or auditory alerts when new events are detected. This is achieved through periodic polling or WebSocket connections, depending on the deployment configuration and latency requirements. Notifications are prioritized based on alert severity, ensuring that critical events receive immediate attention. The interface provides summary dashboards with visualizations of alert trends, top offenders by MAC address, and distribution of alert types over time, supporting situational awareness and informed decision-making by administrators.

\subsubsection{Alert Log}

A feature has been added to allow ordinary users to securely save generated alerts for their own protection and to establish a verifiable audit trail. Through a form on the personal alerts view page, users can register alerts, which generates a JSON file containing alert details alongside a unique cryptographic hash for authenticity. This JSON and hash pairing creates a reliable and tamper-evident record of the alert, securing vital information and enabling users to prove the integrity of the data at a later time. The hash acts as a digital fingerprint for each alert, ensuring its integrity and preventing unauthorized modifications.

The hash generation strategy, detailed in Algorithm~\ref{alg:generateHash}, relies on the SHA-256 hashing algorithm, a cryptographic hash function widely recognized for its collision resistance and computational efficiency. SHA-256 processes the alert information and outputs a fixed-length 256-bit hash value, represented in hexadecimal format for human readability and storage convenience. This approach provides users with authentic proof, enhancing the security of alert information and allowing the alert to serve as verifiable evidence if needed in future situations, such as internal audits, legal proceedings, or compliance reviews. By registering each alert with this method, users gain a secure and trustworthy way to document and reference significant events, reinforcing both the security and reliability of alert information.

The function \texttt{GenerateHash} implements the hash computation in a straightforward and reproducible manner. In line 2, the input string for hashing is created by concatenating the alert's unique identifier \texttt{id} and the monitored endpoint's identifier \texttt{pcId}, forming a composite key that uniquely identifies the alert within the system. This concatenation ensures that even alerts with identical content but originating from different endpoints or at different times will produce distinct hash values. In line 3, the SHA-256 hashing algorithm is initialized using the \texttt{MessageDigest.getInstance("SHA-256")} command, which instantiates a cryptographic digest object configured for SHA-256 computation.

In line 4, the input string is converted to bytes using UTF-8 encoding, ensuring consistent hashing across different platforms, programming languages, and character sets. UTF-8 is chosen for its widespread adoption and compatibility with international character representations. These bytes are then passed to the \texttt{digest} method, which applies the SHA-256 transformation and produces a hashed byte array stored in \texttt{encodedHash}. This byte array represents the raw 256-bit hash output.

The algorithm then proceeds to iterate over each byte in the hashed array (lines 6--12), converting each byte to its hexadecimal string representation. Hexadecimal encoding is used because it provides a compact and human-readable format for binary data. In line 7, each byte is converted to a hexadecimal string using the \texttt{Integer.toHexString(0xff \& b)} operation, where the bitwise AND with \texttt{0xff} ensures that the byte is treated as an unsigned 8-bit value. If the resulting hexadecimal value has only one character (indicating a value less than 16), a leading '0' is prepended in line 9 to ensure a consistent two-character format for each byte. This padding is essential for maintaining the fixed length of the final hash string and for ensuring that the hash can be reliably parsed and compared.

Each hexadecimal representation of the bytes is then concatenated into the \texttt{hexString} variable (line 11). Finally, in line 13, the function returns this \texttt{hexString}, representing the secure SHA-256 hash for the alert. This hash serves as a unique and reliable verification method for alert authenticity, enabling users to detect any tampering or corruption of the alert data by recomputing the hash and comparing it to the stored value. The deterministic nature of SHA-256 ensures that identical inputs will always produce identical outputs, while even the smallest change to the input will result in a completely different hash, providing strong integrity guarantees.

\begin{algorithm*}
\caption{Hash generation function for alert integrity verification.}\label{alg:generateHash}
\scriptsize
\begin{algorithmic}[1]
\Function{GenerateHash}{id, pcId}
    \State input $\gets$ id $\|$ pcId \Comment{Concatenate alert and endpoint identifiers}
    \State digest $\gets$ \texttt{MessageDigest.getInstance("SHA-256")} \Comment{Initialize SHA-256}
    \State encodedHash $\gets$ \texttt{digest.digest(input.getBytes("UTF-8"))} \Comment{Compute hash}
    \State hexString $\gets$ empty string
    \For{\textbf{each} byte $b$ in encodedHash}
        \State hex $\gets$ \texttt{Integer.toHexString(0xff \& b)} \Comment{Convert byte to hex}
        \If{hex.length() == 1}
            \State hexString $\gets$ hexString + '0' \Comment{Pad single-digit hex values}
        \EndIf
        \State hexString $\gets$ hexString + hex
    \EndFor
    \State \Return hexString \Comment{Return 64-character hexadecimal hash}
\EndFunction
\end{algorithmic}
\end{algorithm*}

\subsection{Predict Model}

The solution includes a core module employing prediction models to detect offensive content in Portuguese, Spanish, and English, with three models trained for each language. To ensure computational efficiency, training was set to stop when the loss function reduction became negligible or a maximum number of iterations was reached. This prevented overfitting and non-convergence. 

Error monitoring tracked training and validation losses to identify divergence or overfitting. Early stopping was applied when validation loss increased despite decreasing training loss, ensuring model generalization. Robust optimization techniques further addressed risks like premature convergence or local minima.

While trained models (Logistic Regression, Support Vector Machine, and Multinomial Naive Bayes) handle primary detection, GPT-3 serves as a fallback mechanism, verifying phrases not flagged by the primary models to enhance accuracy. The following sections detail the development process, from dataset selection to model training.

\subsubsection{Choice of Datasets}

To ensure the effectiveness and accuracy of the models, two datasets were selected for each language, resulting in a total of six datasets. These choices were guided by the need for diverse and contextually relevant data, particularly datasets containing real-world language patterns and social media interactions where hate speech and offensive language are commonly encountered.

For Portuguese, the datasets "Linguistic Datasets for Portuguese" \footnote{{Available at https://github.com/EticaAI/linguistic-datasets-portuguese}} \citep{portuguese1} and "BraSNAM2018 Dataset - Análise de sentimentos em tweets em português brasileiro"\footnote{{Available at https://github.com/danielkansaon/BraSNAM2018-Dataset-Analise-de-sentimentos-em-tweets-em-portugues-brasileiro}} {\citep{portuguese2}} were chosen, both accessible on GitHub. These datasets were selected because they include a large volume of user-generated content from Brazilian Portuguese social media interactions, particularly tweets, making them suitable for capturing variations in hate speech expressions across the Portuguese language.

In Spanish, the datasets "Overview of the Task on Automatic Misogyny Identification at IberEval 2018"\footnote{Available at https://amiibereval2018.wordpress.com/important-dates/data/} \citep{spanish1} and "Overview of MEX-A3T at IberEval 2018 - Authorship and aggressiveness analysis in Mexican Spanish tweets"\footnote{Available at https://sites.google.com/view/mex-a3t2018/our-members} \citep{spanish2} were chosen. These datasets are specifically focused on detecting misogynistic and aggressive language in Mexican Spanish tweets, making them highly relevant for training models in the context of identifying hate speech and harmful language within the broader Spanish-speaking social media space.

For English, we selected "Improved Cyberbullying Detection Using Gender Information"\footnote{{Available at https://www.noswearing.com/dictionary}} \citep{english1} and "Are You a Racist or Am I Seeing Things? Annotator Influence on Hate Speech Detection on Twitter"\footnote{{Available at http://github.com/zeerakw/hatespeech}} \citep{english2}. These datasets were chosen for their extensive collection of tweets and user interactions related to cyberbullying and hate speech. They are particularly valuable because they include metadata, such as annotator information, which can influence the detection of hate speech and add depth to model training in complex cases.

The selection of these datasets was driven by their large number of labeled phrases and their origin from real social media environments like Twitter, Facebook, and YouTube. These factors make the data more representative and applicable to the solution’s context, as they capture authentic instances of language used in environments where hate speech and offensive content are prevalent.

\subsubsection{Data Processing}

The data processing pipeline was designed to transform raw user-generated text into standardized numerical representations suitable for supervised hate-speech detection models. The pipeline comprises (i) text normalization, (ii) feature extraction via TF--IDF vectorization, and (iii) sampling and validation strategies to reduce bias and improve generalization.

Text normalization aims to reduce surface variability while preserving semantic cues relevant to classification. Given an input text $x$, a normalized version $\tilde{x}$ is obtained by applying a composition of deterministic transformations,
\begin{equation}
\tilde{x} = f_{\text{norm}}(x) = f_{\text{lem}} \circ f_{\text{stop}} \circ f_{\text{clean}} \circ f_{\text{lower}}(x),
\label{eq:norm_pipeline}
\end{equation}
where $f_{\text{lower}}$ lowercases the text, $f_{\text{clean}}$ removes noise such as URLs, mentions, and non-linguistic symbols, $f_{\text{stop}}$ removes stopwords, and $f_{\text{lem}}$ applies stemming or lemmatization to map tokens to a canonical form. This procedure reduces sparsity in the feature space and improves robustness against superficial orthographic variations.

After normalization, texts are mapped to numerical vectors using TF--IDF. Let $\mathcal{D}=\{d_1,\dots,d_N\}$ be the corpus, and let $t$ denote a term in the vocabulary. The term frequency in document $d$ is defined as
\begin{equation}
\mathrm{tf}(t,d) = \frac{n_{t,d}}{\sum_{t'} n_{t',d}},
\label{eq:tf}
\end{equation}
where $n_{t,d}$ is the number of occurrences of term $t$ in $d$. The inverse document frequency is computed as
\begin{equation}
\mathrm{idf}(t) = \log\left(\frac{N + 1}{\mathrm{df}(t) + 1}\right) + 1,
\label{eq:idf}
\end{equation}
where $\mathrm{df}(t)$ is the number of documents containing $t$. The TF--IDF representation is then
\begin{equation}
\mathrm{tfidf}(t,d) = \mathrm{tf}(t,d)\cdot \mathrm{idf}(t).
\label{eq:tfidf}
\end{equation}
To ensure uniform scaling across samples and mitigate the influence of document length, each document vector $\mathbf{x}_d$ is optionally L2-normalized:
\begin{equation}
\hat{\mathbf{x}}_d = \frac{\mathbf{x}_d}{\lVert \mathbf{x}_d\rVert_2 + \epsilon},
\label{eq:l2norm}
\end{equation}
where $\epsilon$ avoids division by zero. This normalization makes cosine similarity equivalent to the dot product and stabilizes downstream optimization.

To address potential dataset bias and class imbalance, stratified sampling was used when splitting the dataset into training, validation, and test partitions (70\%, 15\%, 15\%). Let $y \in \{1,\dots,C\}$ be the class label. Stratification enforces approximately preserved class priors:
\begin{equation}
\Pr_{\mathcal{D}_{\text{train}}}(y=c) \approx \Pr_{\mathcal{D}}(y=c), \quad
\Pr_{\mathcal{D}_{\text{val}}}(y=c) \approx \Pr_{\mathcal{D}}(y=c) \quad
\label{eq:stratified_split}
\end{equation}
Moreover, augmentation was applied to increase linguistic diversity by generating perturbed variants of each text (e.g., synonym replacement or paraphrasing). Formally, given an augmentation operator $\mathcal{A}(\cdot)$, the augmented dataset becomes
\begin{equation}
\mathcal{D}' = \mathcal{D} \cup \{(\mathcal{A}(x_i), y_i)\}_{i=1}^{N}.
\label{eq:augmentation}
\end{equation}
This step helps reduce overfitting to lexical artifacts and improves robustness to paraphrases commonly observed in real-world user input.

Model robustness was further assessed with 5-fold cross-validation. For a metric $m(\cdot)$ (e.g., macro-F1), the cross-validated estimate is computed as
\begin{equation}
\bar{m} = \frac{1}{K}\sum_{k=1}^{K} m_k, \quad K=5,
\label{eq:cv_mean}
\end{equation}
where $m_k$ is the score on the $k$-th fold.

Table~\ref{tab:preprocessing_steps} summarizes the main preprocessing operations, their goals, and the expected effect on the learning process.

\begin{table}[ht]
\centering
\scriptsize
\caption{Summary of preprocessing and validation steps.}
\label{tab:preprocessing_steps}
\begin{tabular}{p{3.1cm} p{4cm} p{6cm}}
\hline
Step & Operation (input $\rightarrow$ output) & Purpose / expected effect \\
\hline
Cleaning & remove URLs, mentions, non-alphanumeric artifacts & reduce noise; improve signal-to-noise ratio \\
Lowercasing & ``Text'' $\rightarrow$ ``text'' & reduce vocabulary size; reduce sparsity \\
Stopword filtering & remove frequent function words & reduce uninformative terms; focus on discriminative tokens \\
Stemming/Lemmatization & tokens $\rightarrow$ canonical forms & merge morphological variants; improve consistency \\
TF--IDF vectorization & $\tilde{x}\rightarrow \mathbf{x}_d$ using Eqs.~\ref{eq:tf}--\ref{eq:tfidf} & produce weighted features reflecting term relevance \\
Vector normalization & $\mathbf{x}_d \rightarrow \hat{\mathbf{x}}_d$ using Eq.~\ref{eq:l2norm} & stabilize learning; mitigate document length effects \\
Stratified split & preserve class priors (Eq.~\ref{eq:stratified_split}) & reduce sampling bias; fair evaluation \\
Augmentation & $(x,y)\rightarrow(\mathcal{A}(x),y)$ (Eq.~\ref{eq:augmentation}) & increase diversity; improve generalization \\
Cross-validation & mean metric $\bar{m}$ (Eq.~\ref{eq:cv_mean}) & robust performance estimate; reduce variance \\
\hline
\end{tabular}
\end{table}

Overall, these preprocessing steps ensured clean, diverse, and standardized datasets, reducing risks of bias and overfitting, and improving the models' ability to detect hate speech effectively across different linguistic contexts.

\subsubsection{Choice of Models}

Following pre-processing, nine models were selected, with three allocated to each language: Portuguese, Spanish, and English. These languages were chosen for their global significance and the availability of hate speech datasets. The models included Logistic Regression, Support Vector Machine (SVM), and Multinomial Naive Bayes (MNB), selected for their efficiency in text classification and proven performance in hate speech detection \citep{zulqarnain2020comparative}.

Logistic Regression was employed for its strength in multiclass classification using the one-vs-rest (OvR) strategy. This approach enables the model to handle multiple categories by training binary classifiers for each class. Logistic Regression also provides interpretable probabilistic outputs and ensures computational efficiency with high-dimensional data like TF-IDF vectors \citep{hastie2009elements}.

Support Vector Machine was chosen for its ability to handle non-linear data relationships via kernel functions such as the Radial Basis Function (RBF) kernel. Fine-tuning of parameters \(C\) and \(\gamma\) allowed for an optimal balance between margin maximization and misclassification tolerance. This robustness makes SVM particularly effective for the complex class boundaries often encountered in hate speech data \citep{cortes1995support}.

Multinomial Naive Bayes was included for its probabilistic framework and computational efficiency in text-based tasks. By leveraging term frequency distributions, MNB simplifies classification while remaining effective. Laplace smoothing (\(\alpha = 1.0\)) was applied to handle unseen word-class combinations, ensuring reliability even with sparse datasets \citep{brown1992estimating}.

Each model was trained and optimized on language-specific datasets, accounting for linguistic and cultural nuances in Portuguese, Spanish, and English. Preprocessing steps, including tokenization, stemming, and TF-IDF vectorization, ensured consistent feature representation across languages.

The selection of these models reflects their complementary strengths: Logistic Regression as an efficient baseline for linear relationships, SVM for non-linear flexibility, and MNB for rapid performance on sparse datasets. Together, they provide a robust, scalable, and adaptable framework for multilingual hate speech detection.

\subsubsection{Model Training}

To optimize model accuracy and generalization, specific strategies were employed during the training phase using the Scikit-learn library. Each model underwent fine-tuning through grid search optimization to identify the best-performing parameters \citep{MORENO1}. Logistic Regression utilized L1 regularization (Lasso) to automatically select relevant features and prevent overfitting by reducing the weights of less important features to zero. The regularization strength (\(C\)) was set to 1.2, balancing generalization with training data fit. The "saga" solver was chosen for its efficiency with large-scale datasets, and a maximum iteration limit of 10,000 was implemented to ensure convergence.

The Support Vector Machine model employed the radial basis function (RBF) kernel, enabling it to handle non-linear class separations by mapping data into a higher-dimensional space. The regularization parameter (\(C=1.0\)) and kernel coefficient (\(\gamma=0.01\)) were fine-tuned to achieve optimal results. A tolerance value (\(tol=1e-4\)) was used for precise stopping criteria during training, and feature subsets were used during hyperparameter testing to reduce computational overhead.

For Multinomial Naive Bayes, Laplace smoothing (\(\alpha=1.0\)) was applied to avoid zero probabilities for infrequent or unseen words, improving generalization. The \texttt{fitprior} parameter adjusted class priors dynamically based on the observed data distribution, ensuring the model effectively accounted for varying class frequencies.

The datasets were divided into 70\% for training, 15\% for validation, and 15\% for testing. Validation datasets allowed hyperparameter fine-tuning, while test datasets provided unbiased model evaluation. Performance was measured using precision, recall, F1-score, and accuracy to capture each model’s effectiveness in detecting hate speech while minimizing false positives and negatives. Table~\ref{tab:hyperparameters} summarizes the chosen hyperparameters for each model, showcasing the configurations that ensured robust performance across the multilingual datasets.

\begin{table}[h!]
\caption{Hyperparameter Configuration for Each Model}
\label{tab:hyperparameters}
\scriptsize
\centering
\begin{tabular}{lll}
\hline
\textbf{Model} & \textbf{Hyperparameter}         & \textbf{Value}              \\ \hline
Logistic Regression & Regularization (Penalty)       & L1 (Lasso)                  \\ 
                    & Regularization Strength (\(C\)) & 1.2                         \\ 
                    & Solver                         & saga                        \\ 
                    & Maximum Iterations             & 10,000                      \\ 
Support Vector Machine & Kernel                        & RBF \\ 
                    & Regularization (\(C\))         & 1.0                         \\ 
                    & Kernel Coefficient (\(\gamma\)) & 0.01                        \\ 
                    & Tolerance (\(tol\))            & 1e-4                        \\ 
Multinomial Naive Bayes & Laplace Smoothing (\(\alpha\)) & 1.0                         \\ 
                    & Fit Prior (\texttt{fitprior})  & Enabled                     \\ \hline
\end{tabular}
\end{table}

The hyperparameters in Table~\ref{tab:hyperparameters} were chosen empirically through iterative testing on the validation dataset, using grid search and domain-specific knowledge. These configurations ensure that the models generalize effectively to unseen data while maintaining computational efficiency, addressing the challenges posed by high-dimensional and multilingual datasets.

\subsubsection{Use of GPT-3}

The system integrates the GPT-3 API as a supplementary layer to enhance hate speech detection. When the internal models of the Spyware API Gateway fail to classify a phrase as hate speech, GPT-3 is queried as a fallback mechanism. If GPT-3's response differs from the initial prediction, both the phrase and GPT-3's response are saved in a dataset for future retraining, helping refine the internal models over time. This integration is implemented in the function shown in Algorithm \ref{fig:gpt}, which constructs the API request, processes the response, and stores discrepancies for retraining.

In \texttt{VerifyingHateSpeechChatGPT}, the OpenAI API key is set (line 2), and a call to \texttt{openai.Completion.create} (line 3) sends the input text for analysis. The "text-davinci-003" model is used for its high accuracy in text analysis. The \texttt{prompt} dynamically incorporates the input text and requests GPT-3 to identify whether the phrase constitutes hate speech, returning "yes" or "no." The \texttt{temperature} parameter is set to 0.6, balancing response consistency and creativity. The function evaluates GPT-3's response in line 8. If the answer starts with "yes," a log entry is generated (line 9) to indicate hate speech detection, and the function returns \texttt{True}. If the response is "no," the function returns \texttt{False} (line 11). In case of API errors, an error-handling block (line 13) logs the issue (line 14) and defaults to \texttt{False}, ensuring system resilience during connectivity issues.

This approach leverages GPT-3's advanced capabilities to address limitations of traditional models, such as difficulties in identifying rare linguistic patterns or novel contexts. By integrating GPT-3, the system improves detection accuracy in complex cases. Additionally, the storage of divergent predictions for retraining ensures continuous enhancement of the internal models, making the detection system robust and adaptive to evolving linguistic patterns.

\begin{algorithm*}
\caption{Integration with GPT-3}
\scriptsize
\begin{algorithmic}[1]
\Function{VerifyingHateSpeechChatGPT}{text}
    \State \texttt{openai.api\_key} $\gets$ \texttt{API\_KEY\_OPENAI}
    \State response $\gets$ \texttt{openai.Completion.create(}
    \State \ \ \ \ \texttt{model="text-davinci-003",}
    \State \ \ \ \ \texttt{prompt="Identify if this sentence contains hate speech: ''"} + text + \texttt{"''. Answer with yes or no",}
    \State \ \ \ \ \texttt{temperature=0.6)}
    
    \If {response.choices[0].text.lower().find("yes") == 0}
        \State \Return True
    \Else
        \State \Return False
    \EndIf
    
    \If {error occurs}
        \State \texttt{logging("GTP is off!")}
        \State \Return False
    \EndIf
\EndFunction
\end{algorithmic}
\label{fig:gpt}
\end{algorithm*}

\subsection{API Gateway Spyware}

The Spyware API Gateway was developed to allow hate speech detection prediction models to be applied to phrases captured by Spyware, offering a simplified interface. It acts as an intermediary between users and the models, starting with the submission of the phrase to the endpoint. The phrase data frame normalization and vectorization to prepare it for processing by the models. The system identifies the language of the sentence and sends it to the corresponding model. The models, trained on data with and without hate speech, evaluate the presence of hate speech and return the results. This allows users to identify cases of hate speech in the captured sentences practically and efficiently. The API Gateway's centralized structure also guarantees efficient scalability and maintenance.

Language identification is carried out using the "langdetect" library. The API Gateway uses the "TextProcessor" class for text normalization and vectorization. The "TextProcessor" class performs textual pre-processing, including the removal of stopwords and other cleaning steps, standardizing the input text. The normalized and vectorized text is then submitted to the specific model for the identified language. The API Gateway adjusts the input vector to have the size expected by the model, filling empty positions with '0' values if necessary.

The return from the API follows a structured JSON format. When requesting the endpoint, the API Gateway processes the phrase, identifies the language and performs the classification using the corresponding model. The results are stored in a DataFrame called "df\_raw", which contains the original phrase and the classification (1 for hate speech, 0 for not). The DataFrame is converted into a JSON structure using pandas' "to\_json" method, generating a list of JSON objects, each containing the phrase and the rating. This result is returned as a response to the request made to the API Gateway.

\subsection{Classification Pipeline}

The classification pipeline processes textual input to detect hate speech, integrating multiple stages to ensure the preparation and analysis of data are consistent and effective. The pipeline begins with preprocessing, where raw text data is normalized to remove elements such as special characters, hyperlinks, mentions, and stopwords. Tokenization is applied to divide the text into individual components, followed by lemmatization or stemming to reduce words to their base forms, standardizing variations like "running" and "runner" into "run." Text normalization also includes conversion to lowercase, eliminating case sensitivity issues.

After preprocessing, the text is transformed into numerical representations through Term Frequency-Inverse Document Frequency (TF-IDF). This method assigns weights to terms based on their frequency within a document relative to their occurrence across the dataset. Terms frequently associated with hate speech receive higher weights, while common terms such as "the" are downweighted. The resulting vectors are normalized to ensure uniform scaling, which is critical for maintaining balanced contributions across features during classification.

The processed vectors are then evaluated by the prediction models. Logistic Regression analyzes linear relationships between features and the target classes, offering probabilistic outputs that indicate the likelihood of hate speech presence. Multinomial Naive Bayes calculates the likelihood of each class using the distribution of term frequencies, providing a probabilistic approach suited for sparse datasets. Support Vector Machine constructs decision boundaries based on the relationships in feature space, utilizing kernel functions such as the radial basis function to model complex patterns.

The models are trained and evaluated using a parallel processing framework implemented in Python, which allows for efficient handling of input data. Each model operates independently and processes inputs based on the language detected, ensuring adaptability to multilingual data.

The output of the classification pipeline is structured in a consistent format, including the original text, the predicted class (hate speech or non-hate speech), and metadata about the classification process. Any instances of disagreement between the models or ambiguous classifications are flagged for further analysis. These flagged cases are stored for retraining the models, enabling continuous refinement of the system’s performance.

\section{Solution Evaluation} \label{sec_evaluation_aspects}

The evaluation seeks to verify various aspects of the proposed solution, conducting tests for the main components of the architecture:

 \begin{itemize}
   \item API Gateway: covers an analysis of the functionalities offered, the response time of requests, the performance of the hardware, the security implemented and the scalability capacity;
   \item Spyware: evaluating the use of resources on the machine on which it is deployed, the time required to capture data, and the speed of response to requests;
   \item Hate Speech Prediction: evaluate the accuracy of the models used.
 \end{itemize}

The tests not only seek to validate the effectiveness of the individual components but also contribute to an overall understanding of the solution and ensure its viability, security, and performance in the real environment. The tests were conducted on a computer equipped with an Intel i5-3470 processor, 8GB of RAM, and running the Kali Linux operating system. To deploy the application, we chose to use a Docker container, containing a Debian image that includes all the tools needed for the API Gateway to work, including a Redis NoSQL database, a PostgreSQL database, a RabbitMQ messaging service, Prometheus, Grafana, and, finally, the API Gateway SpringBoot application. This approach allowed for the smooth integration of all the tools and functionalities implemented in an isolated and controlled environment, facilitating the configuration and replication of tests in different execution environments.

\subsection{API Gateway}

The evaluation of the API Gateway Central component is essential to ensure its reliability, efficiency, and suitability for managing the complex data flows and communications inherent in the solution. This section focuses on testing the API Gateway’s functionality, response times, and integration with other services, providing insight into its performance under different scenarios. By rigorously examining its capabilities, including endpoint responsiveness and security measures, the goal is to validate the API Gateway as a robust intermediary that handles requests, processes alerts, and manages data in a scalable and secure manner.

\subsubsection{Functionalities}

The API Gateway Central was subjected to extensive functionality testing using Postman to verify endpoint reliability and accuracy. This process involved testing endpoints associated with multiple functionalities, including Alert, Image, Malicious Website, User, Malicious Process, Malicious Port, Bad Language, and Login. For each endpoint, tests were conducted to assess a range of actions: adding new entries, retrieving stored data, deleting records, and validating information.

Both valid, invalid, and edge-case parameter values were tested to evaluate the API’s response under various conditions. This approach ensured that the API could handle typical, erroneous, and extreme input scenarios gracefully. The objective was to confirm that the API functions reliably in different circumstances, providing correct responses and managing errors effectively.

Automated unit and integration tests were also executed before each API Gateway update or deployment, ensuring continuous reliability. All tests concluded successfully, affirming the API Gateway’s capability to handle requests accurately across diverse use cases.

Table \ref{tab:functionality_tests} provides a summary of the tested functionalities, the actions performed, and their success rates, delivering a structured overview of the results. This structured presentation of results helps to quickly understand the functionalities tested and their reliability, reinforcing the API Gateway’s suitability for extensive operational scenarios.

\begin{table}[h!]
\centering
\scriptsize
\caption{Summary of Functionality Tests for API Gateway Central}
\label{tab:functionality_tests}
\begin{tabular}{p{2.5cm}p{4cm}p{5cm}p{2cm}}
\hline
\textbf{Endpoint}        & \textbf{Action}                  & \textbf{Test Parameters}           & \textbf{Result} \\ \hline
Alert                    & Add, Retrieve, Delete, Validate & Valid, Invalid, Edge-case          & Success         \\ 
Image                    & Add, Retrieve, Delete           & Valid, Null, Large File            & Success         \\ 
Malicious Website        & Add, Retrieve                   & Valid, Invalid URL Format          & Success         \\ 
User                     & Register, Login, Delete         & Valid, Invalid Email, SQL Injection & Success         \\ 
Malicious Process        & Add, Retrieve                   & Valid, Non-Existing Process        & Success         \\ 
Malicious Port           & Add, Retrieve                   & Valid, Out-of-Range Port Number    & Success         \\ 
Bad Language             & Add, Retrieve                   & Valid, Special Characters          & Success         \\ 
Login                    & Authenticate                    & Valid, Empty Password, SQL Injection & Success       \\ \hline
\end{tabular}
\end{table}

\subsubsection{Response Time Evaluation}

\begin{figure}[!h]
\caption{Test execution on JMeter}
\label{fig:jmeter}
\centerline{\includegraphics[scale=0.23]{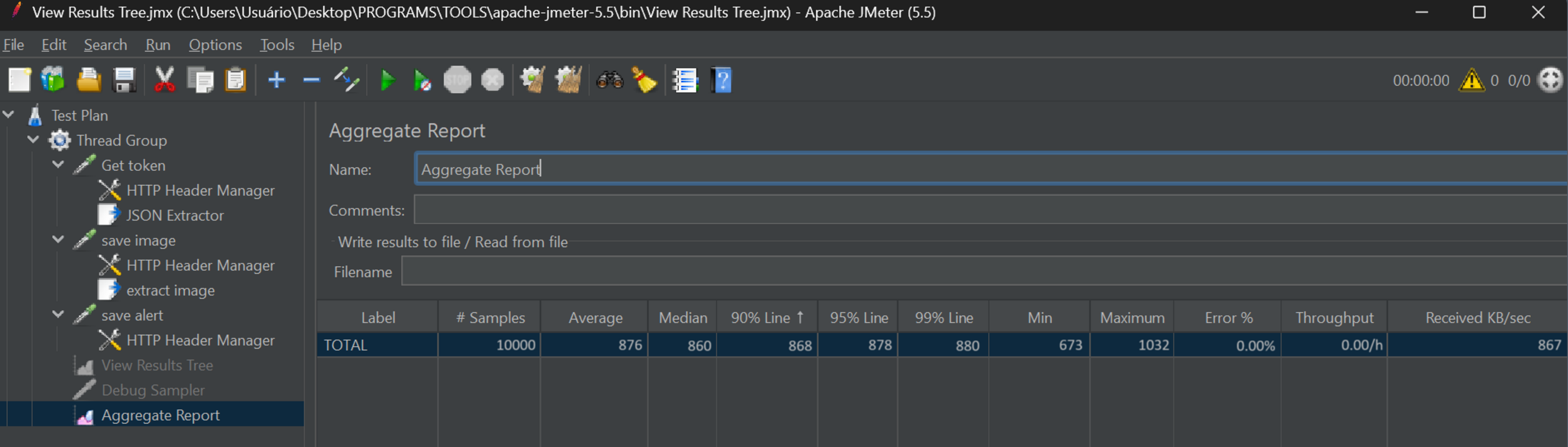}}
\end{figure}

Detailed performance tests were carried out using JMeter, where twenty-five tests were conducted with different configurations of parallel requests to evaluate the performance of API Gateway Central. Each test consisted of three types of requests: authentication, image registration, and alert registration. The requests were executed in varying quantities, such as 500, 1000, 2000, 5000, and 10000 parallel submissions. An example of a test is shown in Figure \ref{fig:jmeter}, which shows the Jmeter screen with a test of 10000 parallel submissions. Due to JMeter's memory limit for sending images in base64 format, a parallel application called API Tester was used for tests with a higher number of requests. The metrics were collected from the Docker image, considering its minimum use of memory and processing.

You can then analyze Table \ref{responsetime}, which shows the average response time for each number of parallel requests.

\begin{table}[!h]
\centering
\scriptsize
\caption{Average Response Time Results}\label{responsetime}
\begin{tabularx}{\textwidth}{lXXXX}
\hline
\textbf{Requests} & 
\textbf{Response Time (ms)} & 
\textbf{Standard Deviation} & 
\textbf{90\% Line (ms)} & 
\textbf{95\% Line (ms)} \\ 
\hline
500 & 112.0 & 6.94 & 132 & 132 \\ 
1000 & 157.5 & 10.29 & 154 & 153 \\ 
2000 & 250.0 & 11.88 & 252 & 250 \\ 
5000 & 471.0 & 29.67 & 330 & 390 \\ 
10000 & 897.5 & 41.58 & 840 & 879 \\ 
\hline
\end{tabularx}
\end{table}

In addition to the response time, it is possible to analyze the standard deviation, the 90\% Line, and the 95\%, which give a more assertive idea of the results obtained, giving a margin of error and being able to predict the most appropriate response time. The 90\% Line shows that 90\% of the samples did not collect more than this time. The remaining 10\% of samples took at least as long. The same for 95\% Line, i.e. 95\% of the samples did not take longer than this time, and 5\% of the remaining samples took at least as long.

Based on works in the literature \citep{chen2001admission}, a maximum response time threshold of 1 second was set. In this way, the response time for requests does not burden the server or compromise the application's functions, making it possible to scale the architecture for use in a real environment. Additionally, comparing with more recent benchmarks, such as the study by Zheng et al. (2010) \citep{5552800}, which evaluated the response time of 5,825 Web services using distributed users, our approach demonstrates competitive performance. Zheng et al. reported an average response time of approximately 1.2 seconds for real-world Web services, validating the relevance of our 1-second threshold as a guideline for ensuring responsiveness while maintaining scalability in distributed environments.


\subsubsection{Performance Evaluation}

For performance, the same tests were carried out, but the hardware metrics were observed. To visualize these results, Table \ref{tab2} was obtained, where it is possible to see the use of memory heap, CPU usage, and how many threads were running, according to the number of parallel requests.

\begin{table}[!h]
\centering
\caption{Average resource usage results\label{tab2}}
\begin{tabular}{cccc}
\hline
\textbf{Requests} & \textbf{Memory Heap (\%)} & \textbf{CPU (\%)} & \textbf{Number of Threads} \\ \hline
500 & 5 & 16 & 22 \\ 
1000 & 8 & 20 & 46 \\ 
2000 & 15 & 24 & 91 \\ 
5000 & 20 & 28 & 120 \\ 
10000 & 39 & 33 & 147 \\ \hline
\end{tabular}
\end{table}

Looking at the results in the table, we can identify significant trends. Firstly, the average use of the memory heap gradually increases as the number of parallel requests increases. For example, for 500 parallel requests, the average memory heap usage is 5\%, while for 10,000 parallel requests, the average usage rises to 39\%.

In addition, we observe an increase in average CPU usage as the number of parallel requests increases. For 500 parallel requests, the average CPU usage is 16\%, while for 10,000 parallel requests, the average usage increases to 33\%.

Finally, the data also shows the average number of threads running for each number of parallel requests. We can see a gradual increase in this number as the application's workload increases. For example, for 500 parallel requests, the average number of threads is 22, while for 10,000 parallel requests, the average number increases to 147.

These results are important for properly sizing the hardware needed to support the desired number of parallel users. Based on the data, it is possible to estimate the memory capacity, processing power, and threads required to meet the expected load. This information aids decision-making to ensure optimized application performance, avoiding resource overload and improving the user experience.

\subsubsection{Scalability Evaluation}

The same load tests were used to measure and predict the application's scalability, applying Linear Regression to model the increase in response time, CPU usage, and memory usage as the number of requests grows.

\begin{figure}[!h] 
\caption{Linear Regression of Response Time, CPU Usage, and Memory Usage} 
\label{fig:reg} 
\centerline{\includegraphics[scale=0.4]{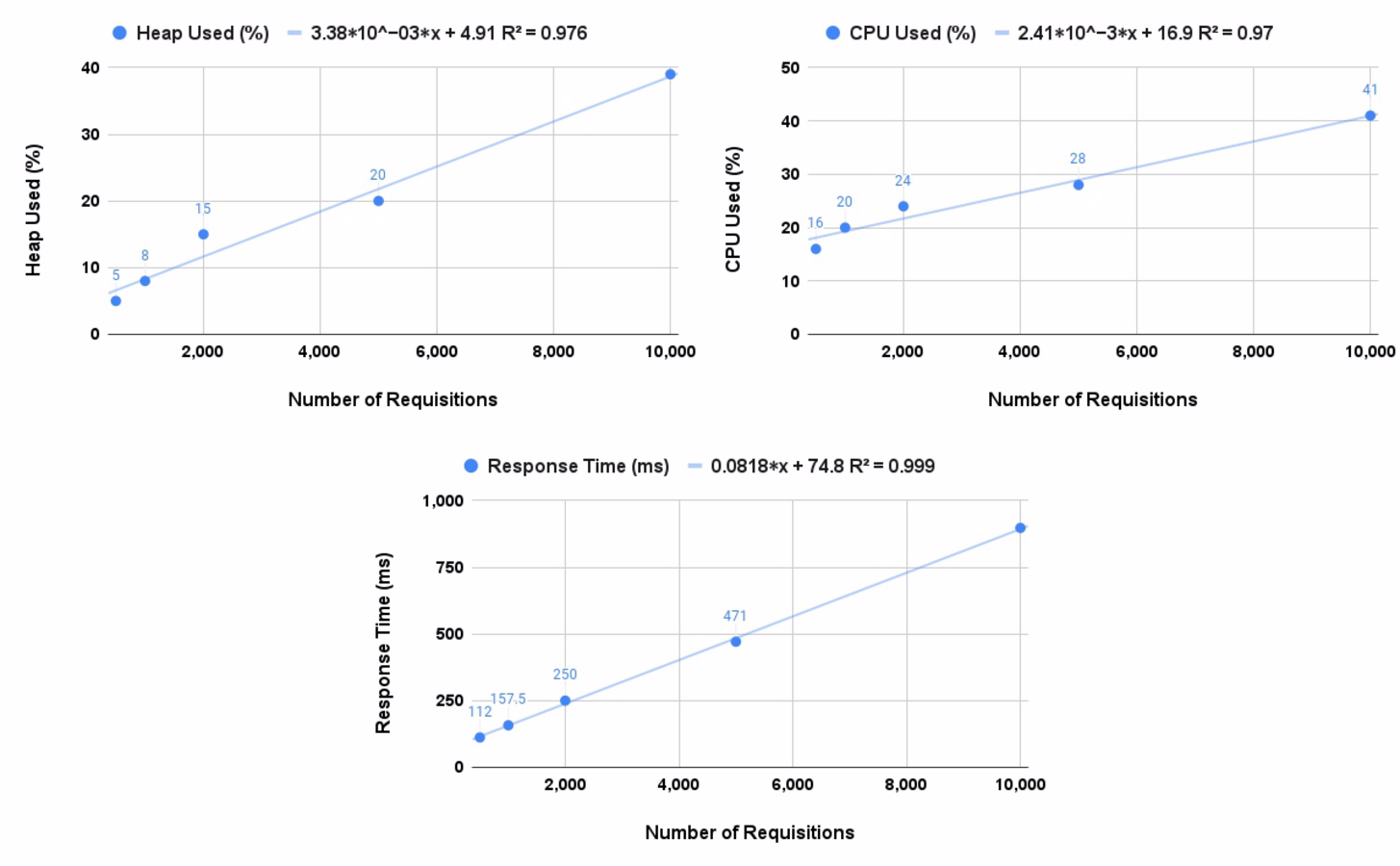}} 
\end{figure}

Linear regression analysis was applied not only to predict response times in real environments by correlating the number of requests with average response time but also to analyze CPU and memory consumption under increasing load. The graphs in Figure \ref{fig:reg} illustrate the trends for response time, CPU usage, and memory heap usage as the number of parallel requests increases. Response time shows an upward trend, represented by the regression formula $74.781 + 0.082x$, which estimates response time based on request volume, enabling predictions of the application's behavior under heavier loads.

Similarly, the CPU and memory usage graphs reveal a direct relationship with the increase in requests, where both metrics grow linearly with the load. The CPU usage is modeled by the regression formula $2.41 \times 10^{-3} \cdot x + 16.9$, indicating that CPU consumption increases steadily with the number of requests. For memory usage, the regression formula $3.38 \times 10^{-3} \cdot x + 4.91$ shows a similar linear trend, illustrating the impact on memory resources as demand rises. These high coefficients of determination across response time, CPU usage, and memory usage indicate a very strong direct linear correlation, providing a reliable basis for performance predictions and resource planning.

Thus, by analyzing Figure \ref{fig:reg}, it is evident that these regression formulas provide valuable insights into resource consumption, allowing us to anticipate the application’s scalability limits and make informed decisions on resource allocation and performance optimizations.

This approach enables anticipation of the performance impact due to an increase in requests, supporting decisions around resource scaling and optimization. Furthermore, the correlation coefficient calculation provides an accurate and mathematically supported result, highlighting the close relationship between resource usage and request volume.

Thus, by analyzing the number of requests, it is possible to observe correlations with various metrics, including response time (with a correlation coefficient of 0.99), memory usage, CPU usage, and thread count, all of which reflect the application's scalability limitations as demand grows.

\subsubsection{Security Assessment}
To ensure the security of the application and prevent vulnerabilities, comprehensive security tests were carried out using an application specially developed to simulate a wide variety of common attacks, such as SQL Injection, Command Injection, XSS Injection, weak passwords, and inadequate data validation \citep{JAVAHERI2024122697}. This test application was built using the Spring Boot framework and consists of a series of HTTP requests that send malicious data to the application's different entry points.

To perform the SQL Injection test, the test application
sends malicious SQL queries as input into form fields or
form fields or URL parameters. The aim is to check that the
application is protected against the execution of these queries and whether it properly filters and escapes the input data to
avoid injection attacks.
Algorithm \ref{alg:postSaveAlert} shows the parameters of a POST request to save an alert. The attempt was to insert a malicious SQL statement into the "processes" field which was intended to delete the "alert" table. The application was initially vulnerable to this type of attack, allowing malicious queries to be executed. However, after this vulnerability was identified, corrections were made to the application to mitigate this threat.

\begin{algorithm}
\caption{Requisition Test with SQL Injection}\label{alg:postSaveAlert}
\scriptsize
\begin{algorithmic}[1]
\State POST \texttt{/alert/save}:
\State \{
\State \ \ \ \ "id": 2,
\State \ \ \ \ "pcId": "string",
\State \ \ \ \ "image": \{
\State \ \ \ \ \ \ \ \ "id": 1
\State \ \ \ \ \},
\State \ \ \ \ "process": "string'; DROP TABLE alert; --",
\State \ \ \ \ "register\_date": "2022-10-25"
\State \}
\end{algorithmic}
\end{algorithm}

Algorithm \ref{alg:postSaveAlertModified}, also for a POST request to save an ''alert'', sought to inject a system command into the ''processes'' input to list the directory. The application initially allowed these commands to be executed, but security measures were implemented to successfully remedy this vulnerability.

\begin{algorithm}
\caption{Requisition Test with Command Injection}\label{alg:postSaveAlertModified}
\scriptsize
\begin{algorithmic}[1]
\State POST \texttt{/alert/save}:
\State \{
\State \ \ \ \ "id": 2,
\State \ \ \ \ "pcId": "string",
\State \ \ \ \ "image": \{
\State \ \ \ \ \ \ \ \ "id": 1
\State \ \ \ \ \},
\State \ \ \ \ "process": "string; ls -la; \#",
\State \ \ \ \ "register\_date": "2022-10-25"
\State \}
\end{algorithmic}
\end{algorithm}

These are just a few examples of the security tests carried out to check that the application prevented attempted attacks, i.e. in these cases, returning "Bad Request". The other security tests followed the same logic and returned a denied request. The testing application covered a wide range of attack scenarios to identify and mitigate potential vulnerabilities. 

\subsection{Spyware Evaluation and Performance Analysis}

The following section presents the evaluation results for the Spyware component, designed to monitor user interactions within the corporate environment. This component focuses on keylogging, screen logging, and process monitoring to detect and mitigate risks related to data leakage and inappropriate behavior. These functions aim to provide insights into user activities that could compromise security, with evaluations assessing the effectiveness, accuracy, and impact on system performance. It is important to note that the performance evaluation was conducted in isolation, focusing exclusively on the resource usage and execution characteristics of the monitoring process itself. The tests were executed on the specified hardware environment detailed earlier to ensure controlled conditions. Metrics such as CPU utilization, memory consumption, and response times were measured to evaluate the impact of the monitoring tasks on the host system.

\begin{figure}[ht]
    \centering
    \includegraphics[width=0.7\linewidth]{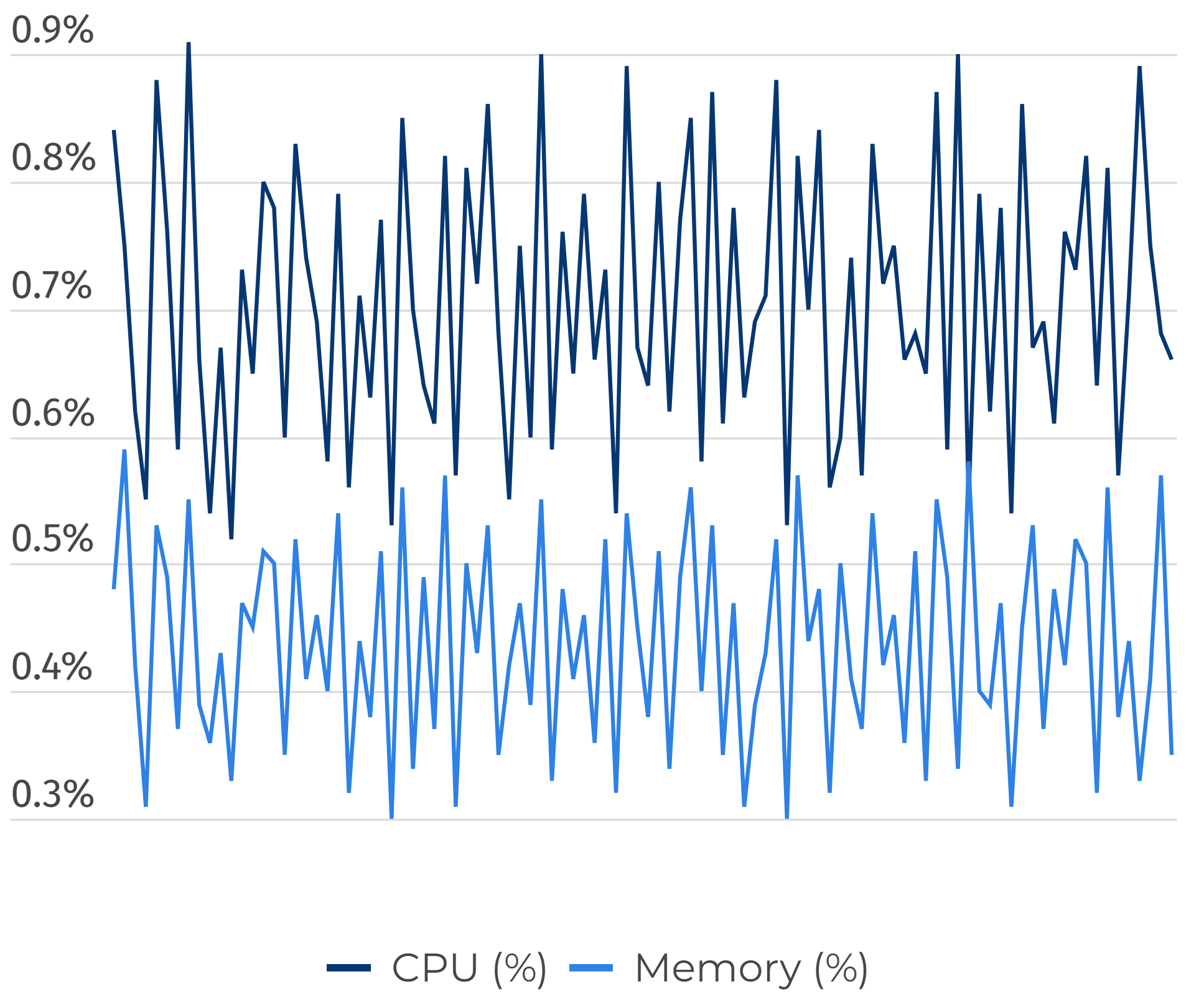}
    \caption{CPU and Memory usage of the Spyware component during a 10-minute execution period.}
    \label{fig:spyware_test_results}
\end{figure}

To evaluate the script’s efficiency, metrics for memory heap, CPU processing, and network usage were recorded over a 10-minute execution period. The average utilization $\bar{U}$ for a resource during the observation window $T$ is defined as:
\begin{equation}
\bar{U} = \frac{1}{T} \int_{0}^{T} u(t) \, dt,
\label{eq:avg_utilization}
\end{equation}
where $u(t)$ represents the instantaneous resource usage at time $t$. Experimental results indicated that the memory heap usage remained consistently below 1\%, indicating stable and minimal resource consumption. Similarly, CPU processing and network usage were both maintained below 1\%, with no significant fluctuations or anomalies observed. As shown in Figure \ref{fig:spyware_test_results}, these metrics confirm the script’s low overhead, ensuring that the monitoring agent does not interfere with the user's primary tasks or system stability.

Beyond resource consumption, the agility of the data capture process was analyzed to determine the latency of alert generation. This process involves the synchronized collection of process lists, screen captures, and telemetry from the Sniffer or KeyLogger threads. The total capture latency $L_{cap}$ can be modeled as the sum of individual task durations:
\begin{equation}
L_{cap} = \max(t_{proc}, t_{screen}, t_{telemetry}) + \delta,
\label{eq:capture_latency}
\end{equation}
where $\delta$ represents the synchronization overhead. Detailed analyses showed that the total capture time averaged approximately 200 milliseconds. This fast response is a result of the optimized Python implementation and the use of non-blocking I/O for evidence gathering. Such efficiency ensures that the system remains responsive, allowing for real-time detection and documentation of suspicious behavior without significant delays that could be exploited by malicious actors.

A critical component of the end-to-end latency is the response time of the Spyware API Gateway, which handles the execution of the compiled prediction models for hate speech detection. The average response time $T_{res}$ obtained after processing the data and applying the inference models was 500 milliseconds. This metric is vital for the application's effectiveness, as it directly affects the speed of feedback and the system's capacity to handle simultaneous requests. While 500 ms is considered satisfactory for the proposed real-time monitoring context, the total system latency $L_{total}$ from event detection to final classification is given by:
\begin{equation}
L_{total} = L_{cap} + L_{net} + T_{res},
\label{eq:total_latency}
\end{equation}
where $L_{net}$ is the network propagation delay. With an average $L_{total} \approx 700$ ms (excluding network jitter), the architecture demonstrates high readiness for production environments.

Table \ref{tab:spyware_performance_summary} summarizes the performance metrics captured during the evaluation phase, highlighting the efficiency of the Spyware component across different operational dimensions.

\begin{table}[ht]
\centering
\caption{Summary of Spyware component performance metrics.}
\scriptsize
\label{tab:spyware_performance_summary}
\begin{tabular}{l c l}
\hline
Metric & Average Value & Evaluation Context \\
\hline
CPU Utilization & $< 1.0\%$ & Sustained monitoring (10 min) \\
Memory Heap Usage & $< 1.0\%$ & Peak load during evidence capture \\
Network Throughput & $< 1.0\%$ & Background policy synchronization \\
Data Capture Latency ($L_{cap}$) & $\sim 200$ ms & Concurrent process and screen logging \\
Inference Response Time ($T_{res}$) & $\sim 500$ ms & Prediction Models API Gateway processing \\
Total Detection Cycle ($L_{total}$) & $\sim 700$ ms & End-to-end flow (local capture to API) \\
\hline
\end{tabular}
\end{table}

In conclusion, the data capture and processing stages were implemented with a focus on high performance and low footprint. The agility in identifying unwanted activity contributes significantly to the safety of the monitored systems. Although the current response times meet the established objectives, continuous monitoring and the potential implementation of edge-side caching or model quantization could further optimize these metrics for scenarios with even higher data volumes or stricter real-time requirements.

\subsection{Prediction of Hate Speech}

This section presents the evaluation of the prediction models used to predict hate speech in english. 

\begin{figure}[htbp]
\caption{Number of 'fold' versus Accuracy}
\label{fig:acuraccy_graph2}
\centerline{\includegraphics[scale=0.27]{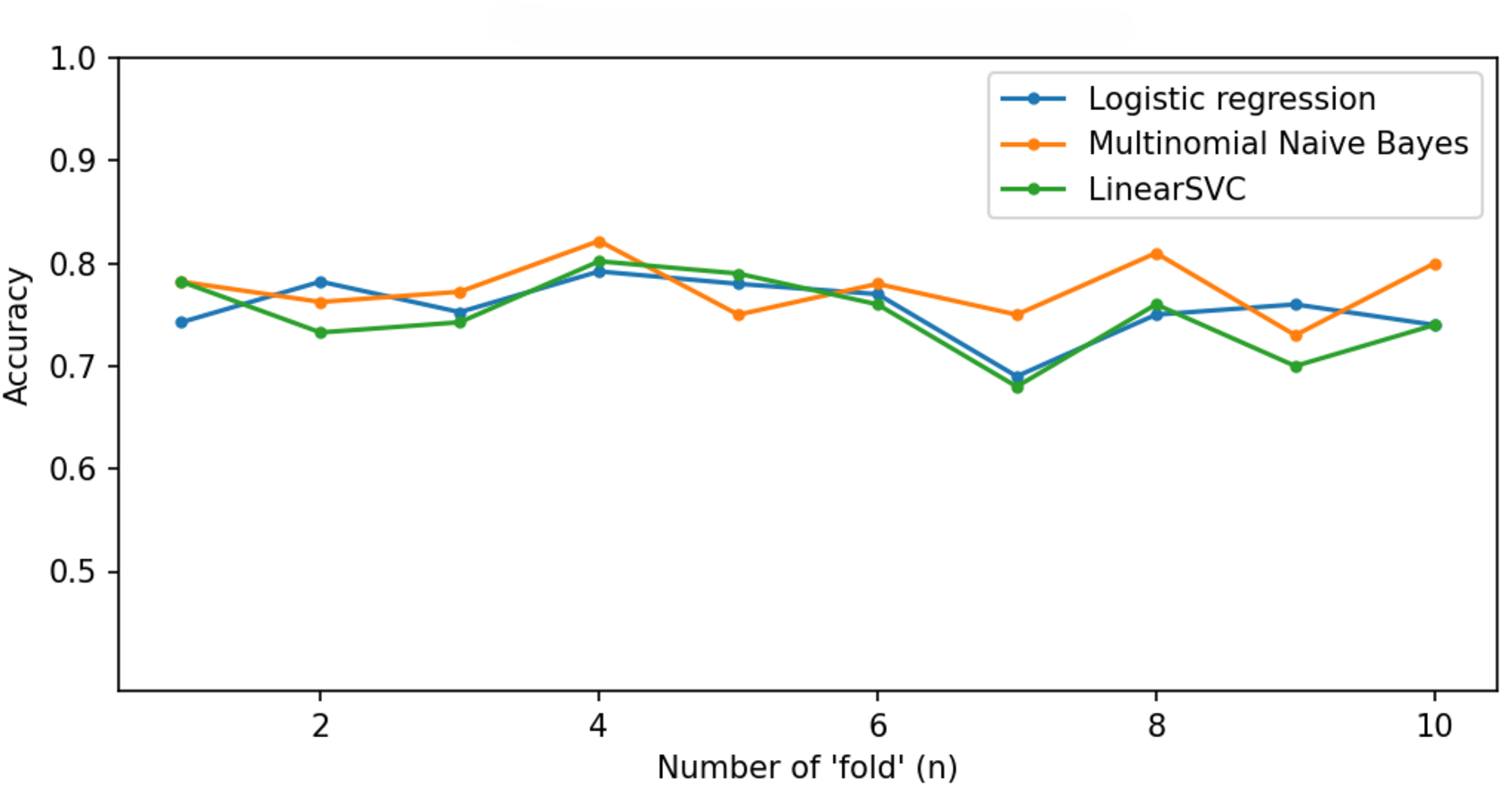}}
\end{figure}

The graph shown in Figure \ref{fig:acuraccy_graph2} displays the performance evaluated by the accuracy metric concerning the number of "folds", which refers to the number of subsets into which the data is divided during cross-validation, for the three models used in the tests. Accuracy is a fundamental measure for evaluating classification models, representing the proportion of correct predictions about the total number of predictions made. The closer it is to 1 (or 100\%), the higher the model's accuracy, indicating a better ability to make accurate predictions.

\begin{figure}[htbp]
\caption{Number of 'fold' versus Balanced Accuracy}
\label{fig:acuraccy_graph1}
\centerline{\includegraphics[scale=0.13]{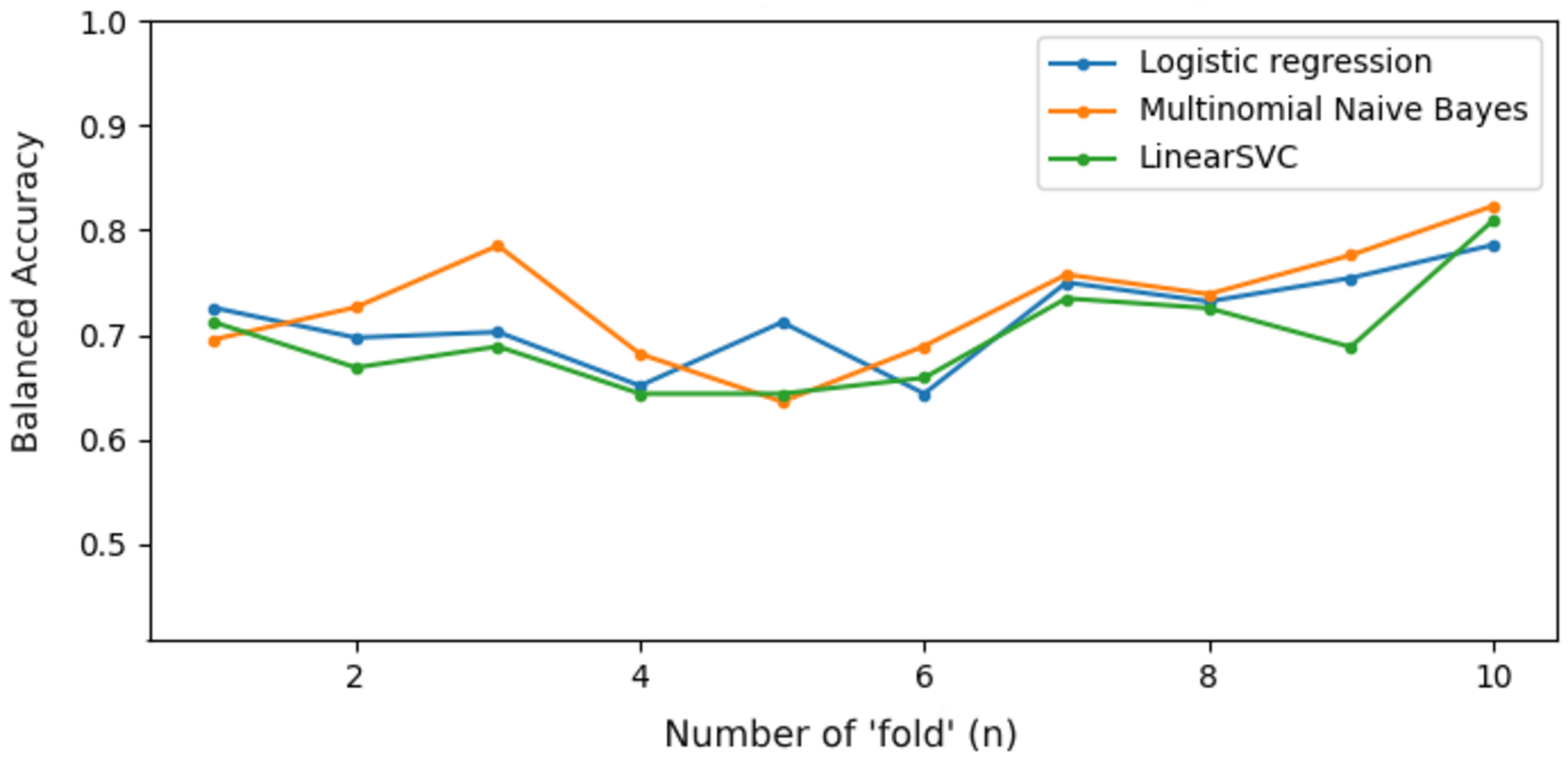}}
\end{figure}

\begin{figure}[htbp]
\caption{Number of 'fold' versus Area Under the ROC Curve}
\label{fig:acuraccy_graph3}
\centerline{\includegraphics[scale=0.13]{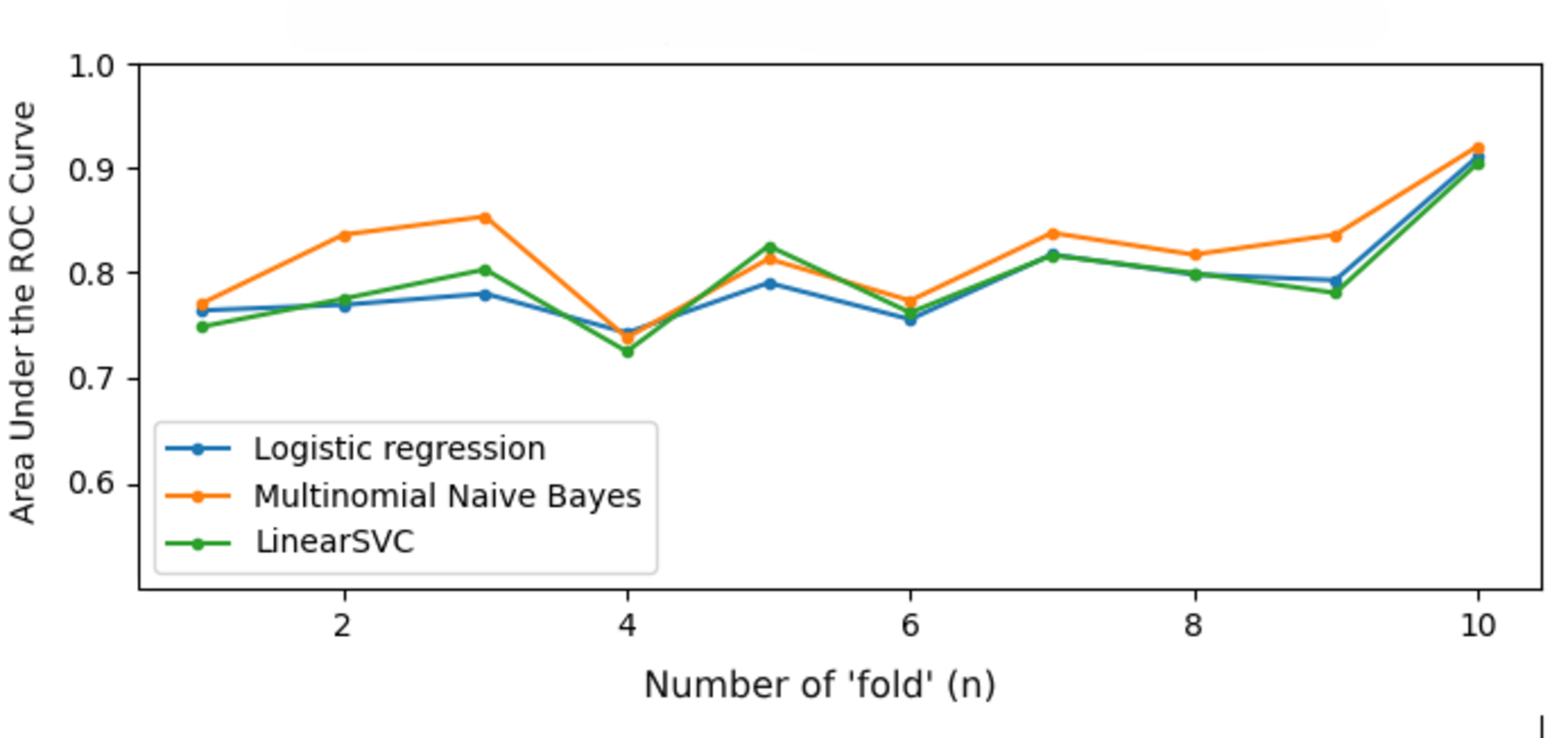}}
\end{figure}

Figures \ref{fig:acuraccy_graph1} and \ref{fig:acuraccy_graph3} show the results of the Balanced Accuracy and the AUC-ROC Curve, respectively, which indicate the effectiveness of predicting hate speech using different strategies. The results for the three models are quite similar, ranging from 75\% to 90\% for the different fold configurations.

It was noted that each of the models performed similarly and acceptably, maintaining an average accuracy of 87\%. However, the Multinomial Naive Bayes model showed a slight advantage over the others. This difference in performance may be related to the specific nature of the dataset and the suitability of the algorithm for the problem in question. The convergence behavior of these models was consistent, as demonstrated by the use of robust optimization techniques, such as the "saga" solver for Logistic Regression and the fine-tuning of \(C\) and \(\gamma\) parameters in the Support Vector Machine. These settings ensured that the models reached an optimal point efficiently without oscillations or divergence.

Even with relatively good performances, the classification models also showed errors, with an average error rate of around 13\%. Error analysis revealed that specific contexts were sometimes misclassified as hate speech, even when they were not. For instance, sentences written entirely in capital letters, which may be used to emphasize or praise, were occasionally flagged as hate speech. Conversely, the models failed to detect hate speech in some cases involving the exaggerated use of special characters or lesser-known offensive terms. These errors highlight the importance of generalization in machine learning models, which was addressed in this study by employing k-fold cross-validation to ensure consistent performance across multiple data splits and by applying techniques such as class balancing and data augmentation.

The results underscore the stability of the models, as evidenced by the consistent metrics across training, validation, and test datasets. Table \ref{tab:results_folds} presents the detailed results for accuracy and AUC metrics across five folds for English, Portuguese, and Spanish datasets, along with the average and standard deviation values. The low standard deviation observed across the folds demonstrates the reliability and consistency of the models across different subsets of the data.

\begin{table}[ht]
\centering
\caption{Cross-validation results for hate speech detection across languages.}
\label{tab:results_folds}
\scriptsize
\begin{tabular}{cccccc}
\hline
\textbf{Fold} & \textbf{EN Acc.} & \textbf{EN AUC} & \textbf{PT Acc.} & \textbf{PT AUC} & \textbf{ES Acc.} \\
\hline
1 & 0.87 & 0.90 & 0.86 & 0.89 & 0.85 \\
2 & 0.86 & 0.89 & 0.85 & 0.88 & 0.86 \\
3 & 0.88 & 0.91 & 0.87 & 0.90 & 0.87 \\
4 & 0.87 & 0.90 & 0.86 & 0.89 & 0.86 \\
5 & 0.86 & 0.89 & 0.85 & 0.88 & 0.85 \\
\hline
\textbf{Mean} & 0.868 & 0.898 & 0.858 & 0.888 & 0.858 \\
\textbf{Std} & 0.007 & 0.008 & 0.008 & 0.008 & 0.008 \\
\hline
\end{tabular}
\end{table}

Regularization techniques, such as L1 regularization in Logistic Regression and Laplace smoothing in Multinomial Naive Bayes, played a significant role in enhancing stability and preventing overfitting, which is crucial for real-world applications. The balanced performance across languages, as highlighted in the table, further supports the generalizability of the models in multilingual scenarios.

These findings highlight the importance of carefully considering context and the nuanced use of words or expressions in hate speech classification tasks. Error analysis and interpretation are critical to improving the models and reducing the error rate. Furthermore, the continuous evaluation and refinement of these models are essential to ensure that they adapt to new data and scenarios, maintaining their effectiveness over time. This includes exploring more advanced architectures, such as deep learning models, in future work to further enhance detection capabilities. Additional test results, detailed analyses, and tests with other languages can be accessed in the project's GitHub repository \citep{spywareAPI}.

\subsection{Evaluation of the Polygon Area Metric}

\begin{figure}[htbp]
\caption{Polygon Area Metric (PAM) for classifier evaluation.}
\label{fig:polygon_pam}
\centerline{\includegraphics[scale=0.4]{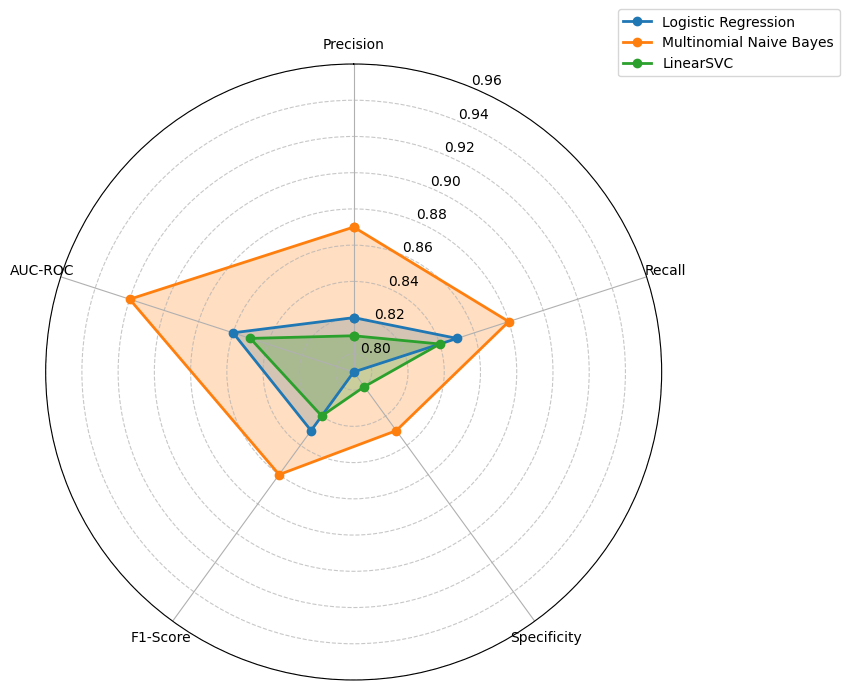}}
\end{figure}

Figure~\ref{fig:polygon_pam} illustrates the evaluation of the classification models using the Polygon Area Metric (PAM), which combines multiple performance indicators into a single geometric visualization \citep{aydemir2020polygon}. Instead of analyzing each metric independently, PAM represents the model as a polygon whose vertices correspond to normalized metric values. In this study, the polygon axes correspond to AUC-ROC, Precision, Recall, F1-Score, and Specificity. The reported results are averaged across the folds of a 5-fold cross-validation procedure, which reduces variance due to the particular train/test split and yields a more stable estimate of performance.

PAM is particularly useful when the goal is to compare models under a multi-objective perspective, because it emphasizes balanced performance. If a classifier performs well on some axes but poorly on others (e.g., high Precision but low Recall), the resulting polygon becomes irregular and the area decreases, revealing trade-offs that may be hidden when reporting only one metric. Formally, if we consider $n$ metrics placed uniformly around a circle and normalized to $r_i \in [0,1]$, the polygon area used by PAM can be computed as:
\begin{equation}
\mathrm{PAM} = \frac{1}{2}\sum_{i=1}^{n} r_i r_{i+1}\sin\left(\frac{2\pi}{n}\right),
\label{eq:pam_area}
\end{equation}
with $r_{n+1}=r_1$. Under this formulation, the area increases when all $r_i$ values are simultaneously high, and decreases when one or more metrics are low, capturing imbalance.

In the context of hate speech detection, the choice of metrics is critical due to the asymmetric cost of errors. Precision measures how reliable positive predictions are:
\begin{equation}
\mathrm{Precision}=\frac{TP}{TP+FP},
\end{equation}
where $TP$ is the number of true positives and $FP$ the number of false positives. High Precision indicates that predicted hate-speech cases are likely correct, which reduces the risk of falsely accusing benign users. Recall measures how many true hate-speech cases were correctly detected:
\begin{equation}
\mathrm{Recall}=\frac{TP}{TP+FN},
\end{equation}
where $FN$ is the number of false negatives. High Recall is important when the goal is to avoid missing harmful content. The F1-Score summarizes the trade-off between Precision and Recall:
\begin{equation}
\mathrm{F1}=2\cdot\frac{\mathrm{Precision}\cdot \mathrm{Recall}}{\mathrm{Precision}+\mathrm{Recall}}.
\end{equation}
Specificity quantifies how well the model avoids incorrectly labeling negative cases as positive:
\begin{equation}
\mathrm{Specificity}=\frac{TN}{TN+FP},
\end{equation}
where $TN$ denotes true negatives. Finally, AUC-ROC captures the model's discriminative ability across all possible decision thresholds, and high AUC-ROC values indicate robust ranking of positive versus negative samples.

The results in Figure~\ref{fig:polygon_pam} indicate that Multinomial Naive Bayes achieved a slightly larger polygon area than the other classifiers. This suggests that its performance was not only strong in a single metric, but also comparatively consistent across the evaluated criteria. In practical terms, a larger PAM implies that the model offers a favorable balance between (i) correctly identifying hate-speech content (high Recall), (ii) avoiding false accusations (high Precision and Specificity), and (iii) maintaining robust separability between classes (high AUC-ROC).

Logistic Regression demonstrated a more uniform polygon shape, which typically reflects stable and well-balanced behavior across metrics. This pattern is often desirable in operational systems because it reduces the likelihood of extreme trade-offs. LinearSVC produced a slightly smaller area, indicating that at least one axis constrained the overall geometric score. Nevertheless, its competitive values in Specificity and F1-Score suggest suitability for deployments where minimizing false positives is prioritized, while still maintaining a strong overall harmonic balance between Precision and Recall.

Overall, PAM complements traditional reporting by providing a single interpretable value (polygon area) and an immediate visual summary of trade-offs. While Multinomial Naive Bayes appears marginally more effective for the evaluated dataset under PAM, Logistic Regression and LinearSVC remain strong alternatives depending on operational constraints. For instance, if the system prioritizes minimizing false positives (to reduce unnecessary investigations), a model with higher Specificity may be preferred even if the PAM is slightly lower; conversely, if the goal is to maximize detection of harmful content, higher Recall may justify the selection of the assurance model, assuming the false-positive rate remains acceptable.

\section{Conclusion} \label{sec_conclusion}

This research developed and validated a comprehensive framework to mitigate data leakage and hate speech in corporate environments by integrating endpoint monitoring techniques with machine learning-based predictive modeling. As organizations increasingly depend on digital infrastructure, the dual challenge of protecting sensitive information and fostering a healthy, inclusive workplace has become paramount. The proposed solution, built upon a modular microservices architecture, demonstrates that it is possible to achieve robust security and behavioral oversight without compromising system integrity or operational performance.

The empirical evaluation confirmed the system's viability across multiple dimensions. The monitoring component (Spyware) exhibited a negligible footprint, maintaining CPU and memory utilization below 1\%, while the data capture and inference pipeline achieved a total end-to-end latency of approximately 700 ms. Furthermore, the hate speech detection models reached an average accuracy of 87\%. The use of the Polygon Area Metric (PAM) provided a holistic view of model performance, revealing that traditional machine learning algorithms, specifically Multinomial Naive Bayes and Logistic Regression, offer a superior balance of precision, recall, and computational efficiency compared to more resource-intensive alternatives.

These findings align with and extend the current literature. While recent studies by Malik et al. (2024) \citep{malik2024deep} and Ramos et al. (2024) \citep{ramos2024comprehensive} highlight the high accuracy of Transformer-based architectures like BERT, our work demonstrates that for real-time corporate monitoring, traditional models provide a more sustainable trade-off between predictive power and hardware requirements. This efficiency is critical for scalability, echoing the quality-of-service considerations discussed by Zheng et al. (2010) \citep{5552800}. By achieving competitive results with lower overhead, our framework proves suitable for deployment in diverse hardware environments, from legacy workstations to modern enterprise systems.

Beyond the technical contributions, this study emphasizes the ethical dimension of workplace monitoring. We argue that such systems must be implemented with transparency, ensuring that employees are informed and that the technology serves as a catalyst for professional development and a positive organizational culture. Rather than a purely punitive tool, the framework should be viewed as a deterrent against harassment and incivility, promoting a safer and more productive environment for all stakeholders.

To advance this research, several directions for future work are proposed:
\begin{itemize}
    \item \textbf{Large-scale Validation:} Transition from controlled environments to production-grade deployments with a significantly larger number of monitored endpoints to assess long-term scalability.
    \item \textbf{Runtime Optimization:} Explore high-performance frameworks such as Quarkus and GraalVM native images to further minimize the footprint of the API Gateway and microservices.
    \item \textbf{Multilingual and Contextual NLP:} Integrate diverse datasets to enhance detection accuracy across multiple languages and investigate the use of quantized Transformer models to capture deeper semantic nuances without excessive resource demands.
    \item \textbf{Privacy-Preserving Monitoring:} Implement advanced anonymization and differential privacy techniques to ensure that monitoring processes comply with evolving global data protection regulations (e.g., GDPR, LGPD).
    \item \textbf{Behavioral Analytics:} Expand the monitoring scope to include cumulative behavioral analysis, identifying patterns of burnout or insider threats through long-term telemetry trends.
\end{itemize}

In summary, the combination of real-time telemetry, multi-objective evaluation through PAM, and a focus on corporate inclusivity ensures that this solution remains a relevant and adaptable response to the evolving technological and ethical challenges of the modern workplace.

\section*{Acknowledgements}

This research was partially funded by national funds through the FCT – Foundation for Science and Technology, I.P. within the scope of projects UIDB/04466/2025 and UIDP/04466/2025. 
This research was also partially funded by national funds through the FCT – Foundation for Science and Technology, I.P. within the scope of project 16881, LISBOA2030-FEDER-00816400, with DOI: \url{https://doi.org/10.54499/2023.16583.ICDT}.

\section*{Disclosure statement}

The authors declare that they have no known competing financial interests or personal relationships that could have appeared to influence the work reported in this paper.

\section*{Funding}

Colaboración Consejería de educación de la Junta de Castilla y León, grupo de investigación ESAL – Expert System and Applications Laboratory (ESALAB).

\section*{Notes on contributors}

\textbf{Darlan Noetzold} contributed to the conception of the study, system architecture design, implementation of the microservices, development of the endpoint monitoring agent, creation of the classification pipeline, data preprocessing, experimentation, and manuscript writing.

\textbf{Anubis Graciela De Moraes Rossetto} contributed to methodological guidance, model validation, experimental design, data interpretation, and critical revision of the manuscript.

\textbf{Juan Francisco De Paz Santana} contributed to the supervision of the research, validation of the predictive models, review of the study’s analytical components, and refinement of the scientific framing and structure of the manuscript.

\textbf{Valderi Reis Quietinho Leithardt} contributed to system modeling, architecture evaluation, performance analysis, interpretation of results, and critical revision of the manuscript, as well as providing domain expertise on pervasive systems and monitoring frameworks.

\end{document}